\documentclass[10pt]{iopart}

\usepackage{fullpage}
\expandafter\let\csname equation*\endcsname\relax
\expandafter\let\csname endequation*\endcsname\relax
\usepackage{mathtools}
\usepackage{amssymb}
\usepackage{amsmath}
\usepackage{amsfonts}
\usepackage{amsbsy}
\usepackage{graphicx}
\usepackage{graphics}
\usepackage{harvard}
\usepackage{iopams}
\usepackage{epstopdf}

\renewcommand{\vec}[1]{\mathbf{#1}}

\DeclarePairedDelimiter{\ev}{\langle}{\rangle}

\begin{document}
\title[Gravitational Waves produced by Compressible MHD Turbulence]{Gravitational Waves produced by Compressible MHD Turbulence from Cosmological Phase Transitions}
\author{
Niksa, Peter$^1$,
Schlederer, Martin$^{1,2}$,
Sigl, G\"unter$^1$}

\address{$^1$II. Institut fuer Theoretische Physik - Universit\"at Hamburg, Luruper Chaussee 149, D-22761, Hamburg, Germany}
\address{$^2$Institut fuer Laserphysik - Universit\"at Hamburg, Luruper Chaussee 149, D-22761, Hamburg, Germany}
\ead{peter.niksa@desy.de, martin.schlederer@uni-hamburg.de, guenter.sigl@desy.de}

\vspace{2pc}

\submitto{\CQG}

\begin{abstract}
We calculate the gravitational wave spectrum produced by magneto-hydrodynamic turbulence in a first order phase transitions.
We focus in particular on the role of decorrelation of incompressible (solenoidal) homogeneous isotropic turbulence, which is dominated by the sweeping effect.
The sweeping effect describes that turbulent decorrelation is primarily due to the small scale eddies being swept with by large scale eddies in a stochastic manner.
This effect reduces the gravitational wave signal produced by incompressible MHD turbulence by around an order of magnitude compared to previous studies.
Additionally, we find a more complicated dependence for the spectral shape of the gravitational wave spectrum on the energy density sourced by solenoidal modes 
(magnetic and kinetic).
The high frequency tail follows either a $k^{-5/3}$ or a $k^{-8/3}$ power law for large and small solenoidal turbulence density parameter, respectively.
Further, magnetic helicity tends to increase the gravitational wave energy at low frequencies.
Moreover, we show how solenoidal modes might impact the gravitational wave spectrum from dilatational modes e.g. sound waves.
We find that solenoidal modes greatly affect the shape of the gravitational wave spectrum due to the sweeping effect on the dilatational modes.
For a high velocity flow, one expects a $k^{-2}$  high frequency tail, due to sweeping.
In contrast, for a low velocity flow and a sound wave dominated flow, we expect a $k^{-3}$ high frequency tail.
If neither of these limiting cases is realized, the gravitational wave spectrum may be a broken power law with index between -2 and -3, extending up to the frequency at which the source is damped by viscous dissipation.
\end{abstract}

\section{Introduction}
The first detection of gravitational waves in 2015 \cite{LIGO2016} opened an exciting new window into observing many different phenomena
invisible in other channels such as black hole mergers and cosmology before photon decoupling.
There are many phenomena like baryogenesis that require beyond the standard model physics.
Also, there are observations which might indicate the presence of large scale (Mpc) magnetic fields in voids \citeaffixed{TVNmin2011}{e.g.}.
A promising avenue to explain the formation of these magnetic fields are cosmological phase transitions \citeaffixed{Vachaspati1991,BBL1996}{e.g.}.
In particular bubbles at a first order electroweak phase transition can produce magneto-hydrodynamic (MHD) turbulence \citeaffixed{Kaminonkowski1994,Sigl1997}{e.g.}.
The resulting initial stochastic magnetic seed fields could then evolve to the magnetic fields that might be present in cosmic voids today.
MHD turbulence produced by a first order electroweak phase transition (EWPT) will inevitably produce gravitational waves 
\citeaffixed{Caprini2002,CDS2008}{e.g.} that might be detectable with gravitational wave detectors such as the planned LISA 
mission\cite{LISA2017}.
In addition, bubble collisions itself produce gravitational waves (GW) that might be observable \citeaffixed{HuberKonstandin2008,CDS2008}{e.g.}, 
depending on the phase transition scenario. 

Similarly, a QCD phase transition (QCDPT), if first order, will also produce gravitational waves and these fall into the frequency range of Pulsar Timing Arrays (PTA)
\cite{Caprini2010} i.e. nHz frequencies.
However, in the nHz frequency range, one expects the gravitational wave signal of supermassive black hole inspirals to dominate the stochastic background  \cite{Sesana2013}.
 
Moreover, an important aspect in producing sufficiently strong magnetic fields on Mpc scales today is magnetic helicity.
Helical magnetic fields affect the evolution of MHD turbulence by driving, under certain conditions, an inverse cascade of energy from smaller to larger length
scales \citeaffixed{Frisch1975,Meneguzzi1981,Cornwall1997,Saveliev2013}{e.g.}.

In the present paper we primarily focus on the gravitational wave signal from MHD turbulence.
\citeasnoun{Kaminonkowski1994} first explicitly showed, using the quadrupole approximation, that bubble collisions in a strong first order phase transition can produce 
turbulence and that the turbulent fluid motion itself can produce gravitational waves.
Their calculations have been extended by \citeasnoun{Kosowsky2002} and \citeasnoun{Dolgov2002}, who went beyond the quadrupole approximation and solved the 
gravitational wave equation directly taking only the Kolmogorov tail into account.
These studies neglected turbulent decay and modeled the turbulence as stationary.
\citeasnoun{Dolgov2002} finds for a turbulent Kolmogorov spectrum a $k^{-7/2}$ scaling for the large frequency gravitational wave power spectrum.
\citeasnoun{Caprini2006} also included the impact of the large scale spectrum into their calculations and showed that for a Kolmogorov spectrum, the GW spectrum 
has three regimes, scaling as $k^{3}$ at large length scales, as $k^{2}$ at intermediary scales and as $k^{-8/3}$ at small scales.
Subsequently, \citeasnoun{Kahn2008a} studied the impact of helical MHD turbulence and showed that it induces polarization into the gravitational wave spectrum.
The most recent semi-analytical evaluation of gravitational waves produced by MHD turbulence by \citeasnoun{CapDurSer2009} has studied the role
of turbulence in more detail,
by studying the evolution from the production of turbulence towards its decay using scaling relations and an assumption regarding the nature of unequal time correlations.
They showed that at small length scales the spectrum should scale as $k^{-5/3}$.
In order to elucidate the spectral shape, \citeasnoun{HindmarshHuber2015} performed relativistic hydrodynamic simulations
in the absence of magnetic fields of the phase transition and 
found that sound waves are a considerable source for gravitational waves. 
The relative fraction of energy in solenoidal (vortical) and dilatational modes (sound waves) depends on the strength of the phase transition \cite{Hindmarsh2017}.
Consequently, for a stronger phase transition more solenoidal motion will be sourced initially, whereas for a weak phase transition more dilatational modes
will be sourced which might later on 
transform into solenoidal motion via either shocks, interactions with 
magnetic fields or solenoidal velocity modes or a non-zero baroclinity $\nabla\rho\times\nabla p$, where $p$ is the pressure and $\rho$ is the density.
The scenarios studied so far have mostly been analyzed over a relatively short time scale after the phase transition and then extrapolated to larger time scales.
Hence they do not provide a complete picture for the solenoidal contribution to the gravitational
wave spectrum and the impact of solenoidal modes on dilatational modes.

One critical factor appearing in the production of gravitational waves via turbulence is the unequal time correlation function (UTC)
of MHD turbulence.
Unequal time correlations are relatively poorly studied and understood in the context of MHD turbulence.
\citeasnoun{CapDurSer2009} tried to apply a model based on a Gaussian function and the local straining time (Lagrangian Eddy turnover time) \cite{Kraichnan1964}, also motivated by \citeasnoun{Kosowsky2002}.
However, it was later discarded since it led to the production of negative energies in the GW power spectrum.
Consequently, \citeasnoun{CapDurSer2009} suggested a top-hat model to imitate the UTC without the appearance of negative energies.
Yet, from a phenomenological perspective that model is poorly motivated.
Here, we improve two important factors with respect to previous studies.
The first factor concerns the evaluation of the UTC $\ev{v(t'),v(t)}$ of a velocity fluctuation at time $t'$ during the sourcing of turbulence and time $t$
during the phase of free decay.
We elucidate this later in more detail, where we discuss a simple model based on random forcing and a general turbulent UTC.
This factor is primarily important for technical reasons in the evaluation and turns out to be critical in making sure that no negative energies end up being evaluated.
The second factor concerns the choice of the UTC, which has also been used by \citeaffixed{Caprini2006,Kahn2008a,Kahn2008b}{e.g.},
is based on a particular timescale.
As we discuss in more detail in section 3.1, the timescale applied in previous studies is in general not applicable here, since it is only adequate for describing temporal 
decorrelation in the frame of comoving trajectories (Lagrangian specification) and the energy transfer, i.e. the shape of the spectrum, rather than that of fixed positions (Eulerian specifications in cosmologically comoving coordinates), which is the frame in which the gravitational wave equation for this problem has been studied.

We begin with a brief recapitulation of the production of gravitational waves from anisotropic stresses and discuss two relevant source terms,
namely magnetic fields and fluid motion in the form of turbulence in the context of isotropic and homogeneous turbulence.
Next, we discuss some key aspects of MHD turbulence and give a detailed discussion of UTCs and clarify the issue of choosing the correct timescale
mentioned above.
Thereafter, we briefly derive the gravitational wave equation for the source terms and the UTC model discussed before.
Also, we present a simplistic scaling analysis for the gravitational wave energy from MHD turbulence.
Furthermore, we briefly discuss some relevant aspects of strongly first order phase transitions.
Finally, we present numerical evaluations of the analytical model and discuss the scaling properties for maximally helical and nonhelical fields and 
discuss how solenoidal modes impact a strongly compressibly driven gravitational wave spectrum.

We use $t$ for physical time and $\tau$ for conformal time.

\section{Generation of gravitational waves by primordial turbulence}
First we briefly review the basic equations that govern the production of gravitational waves by anisotropic stresses in an expanding
flat background universe.
Then, we briefly discuss the stochastic properties of magnetic and kinetic contributions to the anisotropic stress.

\subsection{The gravitational wave equation}
The fundamental equation of interest is the Einstein equation
\begin{equation}
R_{\mu \nu}=8\pi G_{\rm N}\left(T_{\mu \nu}-\frac{1}{2}g_{\mu \nu}T\right), \label{EH_eq}
\end{equation}
where $R_{\mu \nu}$ is the Ricci tensor and $T_{\mu \nu}$ is the energy momentum tensor and $T$ is its trace, while $G_{\rm N}$ is the 
gravitational constant.
For the metric we use the sign convention (-,+,+,+).

The metric is separated into a background and a perturbed part, $g_{\mu \nu}\equiv g_{\mu\nu}^b+h_{\mu\nu}$, where $g_{\mu\nu}^b$ 
is the Friedmann-Robertson-Walker (FRW) metric and $h_{\mu\nu}$ are perturbations of the metric.
In the following, a superscript $h$ denotes a perturbed quantity of $\mathcal{O}(h)$ and a superscript $b$ denotes a background value.
Additionally, we only keep terms of $\mathcal{O}(h)$ in (\ref{EH_eq}).
Analogously, we repeat this separation for the energy momentum tensor
\begin{equation}
T_{\mu\nu}=pg_{\mu \nu}+(p+\rho)u_\mu u_\nu\approx T_{\mu\nu}^b+T_{\mu\nu}^h,
\end{equation}
where $u_\mu$  is the four-velocity with $u^\mu u_\mu=-1$, $T_{\mu\nu}^b$ is the background energy momentum tensor, and $T_{\mu\nu}^h$ is the linearized energy 
momentum tensor of the perturbation. The evolution of the background is governed by the Friedmann equations.

Further, we choose an observer at rest in the FRW metric, such that $u_\mu^b=\left(1,\vec{0}\right)$, $u_0^h=h_{00}/2$ and $u^h_\mu=u_\mu-u_\mu^b$, where $u_\mu^b$ is the background four velocity for a flat spacetime
and $u_\mu^h$ is its perturbation due to the non-flat geometry.
Then, we arrive at $T_{\mu\nu}^b=p^bg_{\mu \nu}+(p^b+\rho^b)\delta_{\mu 0}\delta_{0\nu}$, where $p^b$ is the background pressure and $\rho^b$ is the background 
density of the cosmological fluid and $\delta_{\mu\nu}$ is the Kronecker delta.
$T_{\mu\nu}^h$ is the component of the energy momentum tensor that is of the order of the perturbation.
The perturbed energy momentum tensor can be parameterized as \cite{WeinbergCB2008}
\begin{align}
T_{00}^h&=-\rho^bh_{00}+\rho^h,                          \\
T_{0j}^h&=p^b h_{0j}-(p^b+\rho^b)(\partial_j u^h+u_j^V), 	\\
T_{ij}^h&=p^b h_{ij}+\delta_{ij}a^2 p^h+\partial_i\partial_j\pi^S+\partial_i\pi_j^V+\partial_j\pi_i^V+\pi_{ij}^T,
\end{align}
where the superscript $V$ designates a divergence free vector quantity, $S$ designates a scalar quantity, $T$ denotes a transverse
traceless tensor and $u^h_j=\partial_j u^h+u_j^V$.
Next, we decompose the perturbed metric in a similar manner and we find
\begin{align}
h_{00}^h&=-A,                          \\
h_{0j}^h&=\partial_i B+C_i, 	\\
h_{ij}^h&=P\delta_{ij}+\partial_i\partial_j P^S+\partial_i P_j^V+\partial_j P_i^V+P^T_{ij}.
\end{align}
Consequently, the trace less, divergence free and scalar quantities in the $ij$, $00$ and $i0$ components of equation (\ref{EH_eq}) 
can be grouped into a separate set of equations that have to be fulfilled individually.
For the trace-less tensors we find
\begin{equation}
 \left(\frac{1}{2}\frac{\partial^2}{\partial t^2}-\frac{H}{2}\frac{\partial}{\partial t}-\frac{\ddot{a}}{a}
 -\frac{\nabla^2}{2a^2}\right)P_{ij}^T(\vec{x},t)=8\pi G_{\rm N}\pi_{ij}^T(\vec{x},t), \label{gw-eq1}
 \end{equation}
where $H$ is the Hubble parameter.
Further, the tensorial components are gauge invariant, yet the scalar and vector quantities are not \citeaffixed{WeinbergCB2008}{e.g.}.
Hence, (\ref{gw-eq1}) does describe purely gravitational waves.

However we are not generally interested in an equation for $P_{ij}^T$, but rather we want to calculate $h_{ij}$.
This can be achieved by relating $h_{ij}$ to $P_{ij}^T$ via the quadratic projector
\begin{equation}
 P^2_{ijlm}=P_{il}(\vec{k})P_{jm}(\vec{k})-\frac{1}{2}P_{ij}(\vec{k})P_{lm}(\vec{k}),
\end{equation}
where $P_{jm}(\vec{k})=\delta_{jm}-k_j k_m/k^2$.
Clearly, it is sufficient to show that $T_{ij}$ transforms into $\pi_{ij}^T$ as implied by (\ref{gw-eq1}).
Further, we only consider the impact of anisotropic stresses coming from the flow itself and neglect those coming from perturbations
of the metric within the energy momentum tensor, since those are higher order contributions.
Consequently we subtract the term $p^b h_{ij}$, as this contribution primarily describes the modulation of gravitational waves in vicinity 
of the sound horizon. 
As we argue later on, those scales are less relevant for the GW energy spectrum, 
due to constraints from averaging.
Thus, we find
\begin{equation}
 P^2_{ijlm}\left[T_{lm}^h-p^b h_{lm}\right]=\pi_{ij}^T.
\end{equation}
Continuing, we now rename $P_{ij}^T$ to $h_{ij}$ and $\pi_{ij}^T$ as $\pi_{ij}$.

In the following, we rescale $h_{ij}=a^2 \tilde{h}_{ij}$, such that $g_{ij}=a^2(\delta_{ij}+\tilde{h}_{ij})$ and $\tilde{h}_{ij}$ is 
now a comoving quantity.
This also defines $\tilde{P}_{ij}^T$, which we now rename as $\tilde{h}_{ij}$ and we leave out the tilde from now on i.e. $h_{ij}$ is now a comoving quantity.
Moreover, equation (\ref{gw-eq1}) can be Fourier transformed in $\vec{x}$ and this gives
\begin{equation}
 \left(\partial_t^2+3H\partial_t+\frac{k^2}{a^2}\right)h_{ij}(\vec{k},t)=16\pi \frac{G_{\rm N}}{a^2}\pi_{ij}, \label{gw-eq2}
\end{equation}
where $\vec{k}$ is the comoving wave vector.
Besides, the source term $\pi_{ij}$ is still a physical quantity.
Additionally, we further simplify (\ref{gw-eq2}) by switching to conformal time ${\rm d }\tau={\rm d }t/a$.
Here, and in the following we neglect the time evolution of particle degrees of freedom $g$, since the spectrum is developed
within a timescale that is at most a few Hubble times, hence its change is negligible.
Hence, the gravitational wave equation in conformal time reads
\begin{equation}
\left(\partial_\tau^2+2\mathcal{H}\partial_\tau+k^2\right)h_{ij}(\vec{k},\tau)=16\pi G_{\rm N}\pi_{ij}(\vec{k},\tau), \label{gw-eq3}
\end{equation}
where $\mathcal{H}\equiv(\partial_\tau a)/a= a'/a$ is the conformal Hubble parameter.

Continuing, we define the energy density of the gravitational waves as $\dot{h}^2$ averaged over several wavelengths \cite{MTH-GB1973}
\begin{equation}
 \rho_G(\vec{x},\tau)\equiv\frac{1}{32\pi G_{\rm N}a^2(\tau)}\sum_{ij}\ev{h_{ij}'(\vec{x},\tau)h_{ij}'(\vec{x},\tau)},
\end{equation}
where $\ev{...}$ denotes the spatial average as local GW energy cannot be defined and hence averaging over several wavelengths is necessary.
Expressing the energy via the Fourier transformed strain gives
\begin{equation}
  \rho_G(\vec{x},\tau)=\frac{1}{32\pi G_{\rm N}a^2(\tau)}\int\frac{{\rm d}^3\vec{k}}{(2\pi)^3}\int\frac{{\rm d}^3\vec{k}'}{(2\pi)^3}
  e^{-i(\vec{k}-\vec{k}')\cdot\vec{x}}\sum_{ij}\ev{h_{ij}'(\vec{k},\tau)h_{ij}'(\vec{k}',\tau)}.
\end{equation}
However due to the averaging procedure, only modes with $k\tau\gg1$ are defined in a meaningful way, as near horizon or superhorizon modes
cannot be averaged over several wavelengths and hence do not contribute to causal production of gravitational waves. 

The gravitational wave energy is typically dominated by contributions from smaller scales (large $k$).
Moreover, one important quantity is the power spectrum of gravitational waves
\begin{equation}
 \Omega_{GW}(\tau)=\int{\rm d }\ln(k)\Omega_G(k,\tau) \label{powspec},
\end{equation}
where $\Omega_{GW}(\tau)=\rho_G(\tau)/\rho_C$ is the density parameter of gravitational waves, $\rho_C$ is the critical density and $\Omega_{GW}(k,\tau)$
is the gravitational wave power spectrum.

\subsection{Electromagnetic component}
Next, we specify the electromagnetic component of $\pi_{ij}$, that is $\pi_{ij}^B$.
The electromagnetic energy momentum tensor is 
\begin{equation}
 T_{\mu\nu}^{\rm em}=\frac{1}{4\pi}\left[F_{\mu\alpha}F^{\alpha}_{\nu}-\frac{g_{\mu\nu}}{4}F_{\alpha\beta}F^{\alpha\beta}\right],
\end{equation}
where $F_{\mu\nu}=u_\mu E_\nu-u_\nu E_\mu+\epsilon_{\mu\nu\gamma}B^\gamma$ is the electromagnetic field strength tensor, $B_\mu$ 
is the magnetic four vector, $E_\mu$ is the electric four vector and $\epsilon_{\mu\nu\gamma}\equiv \epsilon_{\mu\nu\gamma\alpha}u^\alpha$ is the Levi-Cevita tensor with 
$\epsilon_{0123}=\sqrt{-\det{g_{\alpha\beta}}}=a^4$ for an FRW metric.
However, we neglect electric fields, as the plasma is highly conducting and uncharged on the scales of interest.
Hence,
\begin{equation}
 T_{\mu\nu}^{\rm em}=\frac{1}{4\pi}\left[B_\gamma B^\gamma(u_\mu u_\nu+g_{\mu\nu}/2)-B_\mu B_\nu\right].
\end{equation}
Further, the magnetic four vector is $B_\mu=a\left(0,\vec{B}^V\right)^T$ and $B^\mu=a^{-1}\left(0,\vec{B}^V\right)^T$, 
where $\vec{B}^V$ is the classical magnetic three vector.
Besides, $\vec{B}^V=\vec{B}^V_0a^{-2}$ where $\vec{B}^V_0$ is the comoving magnetic field vector.
Thus, the spatial Fourier transformed electromagnetic energy momentum tensor becomes
\begin{equation}
 T_{ij}^{\rm em}=\frac{1}{2(2\pi)^4}\int{\rm d}^3\vec{q}\left[B_i(\vec{q},\tau)B_j(\vec{p},\tau)-\frac{1}{2}g_{ij}B_m(\vec{k},\tau)
 B^m(\vec{p},\tau)\right],
\end{equation}
where $\vec{p}=\vec{k}-\vec{q}$.
In conclusion, we fix
\begin{equation}
 \pi_{ij}^B(\vec{k},\tau)=P^2_{ijkl}T_{kl}^{\rm em}(\vec{k},\tau)=\frac{1}{2(2\pi)^4}P^2_{ijkl}\int{\rm d} q^3B_i(\vec{q},\tau)B_j
 (\vec{p},\tau).
\end{equation}
Also, we have neglected the term $h_{\mu\nu}B_\mu B_\nu$, since it is a higher order contribution.
Besides, we define the Alfven velocity $b_i=B_i/\sqrt{4\pi(\rho+p)}$.

Moreover, we define the two point correlation function of Alfven waves by taking an ensemble average and we assume statistical isotropy 
and homogeneity
\begin{equation}
 \ev{b_i(\vec{k},\tau)b_j^*(\vec{q},\tau)}=\frac{(2\pi)^6}{4\pi k^3}\delta(\vec{k}-\vec{q})\left[P_{ij}(\vec{k})E_B(\vec{k},\tau)
 -i\epsilon_{ijl}\frac{k^l}{k}h_B(\vec{k},\tau)\right], \label{mag-cor}
 \end{equation}
 where $E_B$ is the magnetic energy density per $\ln(k)$ normalized to the background density and $h_B$ is a measure for the 
 magnetic helicity density.
 Note, that such an average is only meaningful on sub Hubble scales.
 Explicitly, the magnetic energy density per mode is defined as 
 \begin{equation}
  \rho_B=(\rho+p)\int{\rm d}^3\vec{k} \frac{\ev{\vec{b}(\vec{k})\cdot{\vec{b}(-\vec{k})}}}{2(2\pi)^3}=(\rho+p)\int{\rm d }\ln(k) E_B(k).
 \end{equation}
Further, $h_B$ is normalized such that the case of maximal helicity corresponds to $h_B=\pm E_B$.
Magnetic helicity measures the degree of winding of magnetic field lines and is defined as
\begin{equation}
 H_B=\int{\rm d}^3 \vec{x}\, \vec{A}(\vec{x})\cdot\vec{B}(\vec{x}),
\end{equation}
where $\vec{A}$ is the magnetic vector potential, which in Coulomb gauge is given as $\vec{B}=\nabla\times\vec{A}$.
In Fourier space we find
\begin{equation}
 H_B=\int\frac{{\rm d}^3\vec{k}}{(2\pi)^6}\ev{\vec{A}(\vec{k})\cdot\vec{B}(\vec{-k})}=i\int\frac{{\rm d}^3\vec{k}}{(2\pi)^6 k^2}
 \ev{(\vec{k}\times\vec{B}(\vec{k}))\cdot\vec{B}(-\vec{k})}=\int{\rm d}\ln(k) \frac{(\rho+p)}{k}h_B(k). 
\end{equation}

\subsection{Kinetic component}

Now, we consider the components of $\pi_{ij}$ that come from perturbed fluid motions $\pi_{ij}^V$.
As mentioned before, we are only interested in $T_{ij}-p^b h_{ij}$.
The spatial components are $T_{ij}-p^b h_{ij}=(p+\rho)u_\mu u_\nu$.
As before we apply the projection operator $P^2_{ijkl}$ and Fourier transforming gives
\begin{equation}
 \pi_{ij}^V(\vec{k},\tau)=\frac{1}{(2\pi)^3}(\rho+p)P^2_{ijkl}\int{\rm d}^3\vec{q}\,
 v_i(\vec{q},\tau)v_j(\vec{p},\tau) \label{velcor}.
\end{equation}
Next, we define the kinetic two point function where we assume that the flow is statistically isotropic and homogeneous
\begin{align}
 &\ev{v_i(\vec{k},\tau)v_j^*(\vec{q},\tau)}=\frac{(2\pi)^6}{4\pi k^3}\delta^3(\vec{k}-\vec{q})\Bigl[\left(\lambda\delta_{ij}-\frac{k_i k_j}{k^2}\right)E_S(\vec{k},\tau)+ \nonumber\\
 &2\frac{k_i k_j}{k^2}E_D(\vec{k},\tau)-i\epsilon_{ijl}\frac{k^l}{k}h_V(\vec{k},\tau)\Bigr], \label{kin-cor}
\end{align}
where $E_S$ is the kinetic energy of solenoidal modes and $E_D$ is the kinetic energy of dilatational modes with total kinetic energy $E_V=E_S+E_D$.
The kinetic helicity is $h_V$ and it is normalized such that the maximally helical
fluid is given by $h_V=\pm E_S$.
The difference in the tensorial prefactor of solenoidal and dilatational modes is due to the solenoidal constraint $k_i v^S_i=0$ and due to the dilatational constraint $\epsilon_{ijn} k_j v^D_n=0$.
For the dilatational modes there is no antisymmetric component and $E_S\to E_D$.

 Explicitly, the energy density per $\ln(k)$ is defined as 
 \begin{equation}
  \rho_V=(\rho+p)\int{\rm d}^3\vec{k} \frac{\ev{\vec{v}(\vec{k})\cdot{\vec{v}(-\vec{k})}}}{2(2\pi)^6}=(\rho+p)\int{\rm d }\ln(k) E_V(k).
  \end{equation}
  Kinetic helicity measures the degree of winding of magnetic field lines and is defined as
\begin{equation}
 H_V=\int{\rm d}^3\vec{x} (\nabla\times\vec{v}(\vec{x}))\cdot\vec{v}(\vec{x}).
\end{equation}
As before, in Fourier space one finds
\begin{equation}
 H_V=i(\rho+p)\int\frac{{\rm d}^3\vec{k}}{(2\pi)^6}
 \ev{(\vec{k}\times\vec{v}(\vec{k}))\cdot\vec{v}(-\vec{k})}=(\rho+p)\int{\rm d}\ln(k) k\,h_V(k). 
\end{equation}
In conclusion, we now set $\pi_{ij}=\pi_{ij}^B+\pi_{ij}^V$.

\section{MHD turbulence}

Here we outline MHD turbulence in a simplified picture.
We assume only non-relativistic subsonic velocities with $v\ll c_s=1/\sqrt{3}$ and $v_{A}^2=B^2/(4\pi(\rho+p))\ll c_s$,
such that the assumption of an weakly compressible flow is reasonable.
MHD turbulence is generally described by two Reynold's numbers $Re=v_L L/\nu$ and $Rm=v_L L/\eta$, where
$v_L$ is a characteristic velocity at a length scale $L$, $\nu$ is the viscosity and $\eta=1/(4\pi\sigma\epsilon_0)$ is the magnetic resistivity, $\sigma$ is the conductivity and $\epsilon_0$ is the vacuum permittivity.
In the early universe close to the horizon scale and with velocities close to the speed of light, these numbers are e.g. at 
around the electroweak scale, not necessarily during the phase transition itself, $Re\sim\mathcal{O}(10^{11})$ and $Rm\sim\mathcal{O}(10^{17})$ \cite{ReynoldsEU}.
Consequently the scale of viscous dissipation is far larger than the scale resistive dissipation for magnetic fields, which is generally case for the evolution of MHD turbulence at large scales in the universe.
Turbulence is typically described by two important scales, the injection or integral scale and the dissipation scale.
The dissipation scale is typically the scale for which $Re_d=1=v_d L_d/\nu$ and at the integral scale one has $Re_I=v_I L_I/\nu$.
If the turbulence is fully developed i.e. in a state of self-similar evolution, then the spectrum has a particular scaling between the inertial and the dissipation scale, that is the Kolmogorov spectrum $L^{2/3}$ and the velocity is given by $v=C_K^{1/3}\epsilon^{1/3}L^{1/3}$, where $C_K\sim1.4\to1.7$ is the Kolmogorov constant and $\epsilon$ is the energy dissipation rate, then we find $Re_I\sim (L_I/L_d)^(4/3)$.
Consequently for $Re=10^{11}$ we expect fully developed turbulence to stretch over $8$ orders of magnitude in wave number around the electroweak phase transition.

In general the detailed picture of the evolution of MHD turbulence in the early universe is far more complicated and not required here.
This is, as we see later, due to the fact that the significant window for the production of gravitational waves is around the phase transition itself, even if the source terms itself are long-lasting.

Regardless, at some point in the early universe turbulence will be damped away due to several important viscous phases in the radiation dominated universe.
In particular neutrino and photon diffusion will damp fluid motion away.
At first dissipation of turbulent energy is dominated by neutrinos at $T\sim80\to20$MeV and then by photons at around $T\lesssim 1$MeV \cite{BanJed2004,JedSigl2011}.
In between the photon and neutrino damping regime, one can expect freely evolving turbulence.
The viscous decay will only fully damp away the fluid flow, but not the magnetic fields.
Consequently, between the neutrino and photon diffusion phase, magnetic fields can re-excite fluid motion and hence turbulence. 
The photon damping regime is of importance at $T\lesssim1$MeV, since $e^+ e^-$ pair production ceases and the free electron density drops by roughly $9$ orders in magnitude, which drives a sharp increase of the photon mean free path by 9 orders in magnitude.
Later on turbulence might reappear at around photon decoupling, due to the still preserved magnetic fields.

In summary the above picture tells us that in the radiation dominated phase, turbulence will only be an issue up to the photon damping regime at $T\lesssim 1$MeV and afterwards in the matter dominated regime, which we do not discuss here.
Typically we expect that only a short time integration is relevant and an understanding of the full evolution of the turbulent flow in the early universe is not needed for the discussion here.
This is primarily due to turbulent decay.

Next, we discuss the spectra of a turbulent flow and its scaling properties with time.

\subsection{Spectra and Scaling Properties}

We assume that the kinetic and magnetic energy spectrum are comparable and that both spectra behave according to a Kolmogorov 
inertial range scaling of $k^{-2/3}$ for $k\gg k_I=2\pi/L_I$ and $k<k_d$, where $k_d=2\pi/\lambda$ and $\lambda$ is the dissipation scale, 
which denotes the scale at which turbulent energy is dissipated and $L_I$ is the integral scale.
Further, we assume a Batchelor causal tail $k^{5}$ for $k\ll k_I$.
Generally, these tails occur  in the spectrum if the fluctuations are homogeneous and isotropic and is also required by causality \cite{JedSigl2011}.
At scales $k>k_d$, we assume for simplicity a hard cutoff $E(k>k_d)=0$.
In general the cutoff is exponential, yet since we have a large inertial range, as discussed before, the precise modeling of the cutoff
is expected to have no relevant impact on the results.

Therefore, we apply the von Karman model \cite{VKM1948}
\begin{equation}
 E(k)=C_E \frac{K^5}{(c+K^2)^{17/6}}\theta(L_I/\lambda-K), \label{vKm}
\end{equation}
where $K=kL_I/(2\pi)$ and $c=5/12$ corresponds to $E(k)/k$ having a maximum at $K=1$.
The model above shows in the limit $K\ll1$ the above stated $k^5$ tail and in the limit $k\gg1$ a Kolmogorov spectrum $k^{-2/3}=k^{5-17/3}$ is appearing.

The factor $C_E$ is fixed by the following normalization conditions
\begin{equation}
 \frac{3}{2}\frac{\Omega_t}{\Omega_r}=\int_{-\infty}^\infty{\rm d }\ln(K)E(K), 
\end{equation}
where $\Omega_t$ is the density parameter of either the kinetic or magnetic component of MHD turbulence.
Hence,
\begin{equation}
 C_E=-\frac{3}{2\pi^{3/2}}\left(\frac{10}{3}\right)^{1/3}\frac{\Gamma(17/6)}{\Gamma(-2/3)}\approx 0.172\frac{\Omega_t}{\Omega_r},
\end{equation}
where we assumed $L_I\gg\lambda$ and hence neglected the cutoff scale in the normalization.
We assume that the fields evolve in equipartition, thus $E_S\approx E_B$.

If we neglect the back-reaction from gravitational waves onto kinetic or magnetic fluctuations, since these are higher order contributions, 
we can assume that the evolution of the spectra follows free decay via MHD turbulence.

Due to the likely existence of a universal scaling law in the causal tail \cite{LesieurTFB2008}, we require 
$v^2 L_I(\tau)^5\approx {\rm const}.$ (Loitsyansky constant).
Therefore, we assume that the energy and the integral scale follow a self-similar evolution 
\begin{equation}
 L_I(\tau)=\begin{cases}
            L_*,\quad \tau\leq \tau_0+\tau_{\rm b} \\
L_*\left(\frac{\tau-\tau_0-\tau_b+\tau_D}{\tau_D}\right)^{\gamma_1}, \tau>\tau_0+\tau_{\rm b} \label{levo}
\end{cases}
\end{equation}
where $\tau_D$ is a decay time constant and we set
\begin{equation}
\tau_D=\frac{L_*}{2\sqrt{v_{1,i}^2}}=\frac{L_*}{\sqrt{2\frac{\Omega_{t,*}}{\Omega_r}}},
\end{equation}
where $v_{1,i}$ is a characteristic velocity, which we discuss in the next chapter in more detail.
Subsequently, we assume for the energy the following temporal evolution 
\begin{equation}
 \Omega_t(\tau)=\Omega_{t,*}(\tau_0+\tau_{b})\begin{cases}
         1-\left(\frac{\tau_{\rm b}-(\tau-\tau_0)}{\tau_{\rm b}}\right)^\alpha,\quad  \tau_0\leq \tau\leq \tau_0+\tau_{\rm b}\\
         \left(\frac{\tau_D}{\tau-\tau_{b}+\tau_D}\right)^{\gamma_2}, \quad \tau\geq\tau_{\rm b}+\tau_0, \label{enevo}
                                                          \end{cases}
\end{equation}
where $\tau_{\rm b}$ is the time scale over which the initial turbulent spectrum is generated, the power law index $\alpha$ can be used to model a nonlinear build up time 
and we generally assume $\tau_{\rm b}=\tau_D$.
Such an evolution corresponds to an turbulent cascade, in which the energy is being transported to smaller scales towards the dissipation scale and the integral scale shifts to larger scales.
Here, $\Omega_{t,*}$ is the initial turbulent energy density and $L_*$ is the initial integral scale at the phase transition.

We assume from now on that the turbulent spectrum is produced linear in time, i.e. $\alpha=1$.
For the exponent $\gamma_1$ and $\gamma_2$ we use \citeaffixed{LesieurSchertzer1978,Olesen1997}{e.g.}
\begin{equation}
 \gamma_2=2\frac{\sigma+1}{\sigma+3},\quad \gamma_1=\frac{2}{\sigma+3},
\end{equation}
where $\sigma+1$ is the scaling of the large scale tail.
For $\sigma=4$ we have $\gamma_{1}=2/7$ and $\gamma_{2}=10/7$.
\citeasnoun{Brandenburg_DecClasses} argue that for MHD turbulence the dependence is more difficult, since $\sigma$ will not be representative of the large scale tail alone,
rather $\sigma$ is representative for the relevant invariants (e.g. like the above Loitsyansky constant or magnetic helicity).
They discuss four different scenarios $\sigma=4,2,1,0$.
The case $\sigma=2$ generally describes non-helical magnetically (or at least equipartition) driven turbulence and hence $\gamma_{2}=6/5$ and$\gamma_{1}=2/5$.
If some moderate magnetic helicity is present, e.g. $h_B\sim 0.1$, a scenario with $\sigma=1$ is appropriate and for maximally 
helical MHD turbulence $\sigma=0$ is expected which gives $\gamma_{1}=\gamma_2=2/3$.
The scenario $\sigma=4$ is representative of pure hydrodynamic decay, yet likely not relevant for MHD turbulence.

The scaling for the case with magnetic helicity is indicative of an inverse cascade.
In that case, the energy dissipation rate is slower, while the integral 
scale grows faster i.e. $\Omega_T(\tau)L(\tau)=const.$. 
In contrast, for the $\sigma=4$ case the energy spectrum evolves along the initial large scale tail.
We will primarily focus on the cases $\sigma=0$ (maximal helical) and $\sigma=2$ (non-helical).

For compressible MHD turbulence we also study a slightly different scenario.
We assume a $k^{-2}$ spectrum, if the turbulence is driven by dilatational (purely compressible) modes \citeaffixed{Sun2017}{e.g.},
\begin{equation}
 E_D(k)=C_D \frac{K^5}{(c_D+K^2)^{3}}\theta(L_I/\lambda-K), \label{vKmS}
\end{equation}
where $c_D=1/2$ and 
\begin{equation}
 C_D=\frac{9}{32\sqrt{2}}\frac{\Omega_t}{\Omega_r}\approx0.2\frac{\Omega_t}{\Omega_r}.
\end{equation}
For the velocity spectrum we then have
\begin{equation}
 E_V(k)=f_DC_D \frac{K^5}{(c_D+K^2)^{3}}\theta(L_I/\lambda-K)+f_SC_E \frac{K^5}{(c+K^2)^{17/6}}\theta(L_I/\lambda-K),
\end{equation}
where $f_D$ and $f_S$ denote the fraction of dilatational and solenoidal modes, respectively, with $f_D+f_S=1$.
Note that cross correlations between dilatational and solenoidal modes vanish, as their spectral tensor is 0.
Typically for subsonic flows i.e. $\Omega_V/\Omega_r<2/9$ one can assume that $f_S>f_D$, whereas for supersonic flows $f_D>f_S$.
In general, the initial ratio depends on the source of the fluctuations i.e. the relevant forcing. 
For the temporal decay, we assume the same behavior as for the incompressible case discussed before i.e. for nonhelical turbulence
where $\gamma_2=6/5$ and $\gamma_1=2/5$.

Furthermore, for the temporal evolution of the dilatational mode fraction $f_D$ we consider several models.
First, our model A for the evolution of $f_D$ is
\begin{equation}
 f_D(\tau)=f_{D,i}\begin{cases}
            1-\frac{\tau-\tau_0}{\tau_D}, \quad \tau<\tau_0+\tau_D\\
            0, \qquad \tau>\tau_0+\tau_D, \label{modA}
           \end{cases}
\end{equation}
i.e. dilatational modes are converted into solenoidal modes on a timescale of $\tau_D$.
Here, $f_{D,i}$ is the initial fraction of dilatational modes.
Next, in model B we assume that the fraction of solenoidal modes remains constant
\begin{equation}
 f_D(\tau)=f_{D,i} \label{modB}.
\end{equation}
Last, in model C we assume that the fraction of dilatational modes remains constant during the production of fluid motion and is then 
converted into solenoidal modes over a time $s\tau_D$,
\begin{equation}
  f_D(\tau)=f_{D,i}\begin{cases}
	    1, \qquad \tau<\tau_b+\tau_0\\
            1-\frac{\tau-\tau_0-\tau_b}{s\tau_D}, \quad \tau_0+\tau_b<\tau<\tau_0+s\tau_D+\tau_b\\           
            0, \qquad \tau>\tau_0+s\tau_D+\tau_b \label{modC}.
           \end{cases}
\end{equation}
For the parameter $s$ we use $s=1$ and $s=2$, denoted as model C$_1$ and C$_2$.
These parameterizations bracket reasonable possible evolutions.

\subsection{Unequal Time Correlations}

Now we discuss unequal time correlations (UTC), also called temporal decorrelation, by studying some concepts in more detail.
For a more-in-depth, but brief, and recent review on turbulent UTCs we refer the reader to \cite{HeJinYang2017}.
In equation (\ref{mag-cor}) we only considered equal time-correlation, but here we require knowledge about correlations 
between fluctuations at different times as we will discuss later on.
Hence, we generalize equation (\ref{mag-cor}) and equation (\ref{kin-cor}) by some additional factor that depends on the time 
difference.
In the following, we briefly discuss a few different models for UTCs.
First we primarily focus on the purely incompressible case.

\subsubsection{Decorrelation of freely decaying incompressible fluctuations}\hspace*{\fill}
\newline
One model developed by \possessivecite{Kraichnan1964} for Lagrangian fluctuations in order to understand turbulent energy transfer and studied by \citeasnoun{Gogoberidze2007} and 
\citeasnoun{CapDurSer2009} is to model decorrelation via a Gaussian function as 
\begin{equation}
\ev{v(t,k)v(t',k)}=\ev{v(t,k)v(t,k)}\exp\left[-\frac{1}{2}\left(\frac{t-t'}{t_L(k)}\right)^2\right],
\end{equation}
where $t_L\propto k^{-2/3}$ is the local straining time (also known as Lagrangian eddy turnover time).
A typical estimate for the timescale is $t_L(k)\sim 1/\sqrt{k^2E(k)}$.
Typically an integrated formulation of the timescale is considered \cite{Pouquet1976}
\begin{equation}
 t_{L}^{-1}(k)\approx 0.3\sqrt{\int_0^k qE(q) {\rm d}q}. \label{lagtime}
\end{equation}
Such a timescale describes local interactions.
This timescale is also called Lagrangian eddy turnover time or local straining time and describes the UTC in the reference frame of a given fluid particle as it is primarily driven by the local flow.
The Lagrangian timescale is critical for the description of energy transfer properties of the flow and hence drives the appearance of a Kolmogorov spectrum \cite{Pouquet1976}.
Since turbulence is driven by non-local interactions e.g. small scale fluctuations are driven by large scale fluctuations, one can expect that decorrelation might not only be described by local interactions.

Here, we require the Eulerian eddy turnover time, also known as sweeping time, since we study the UTC of the field, rather than the trajectories.
Under the term Lagrangian velocities $v_L$, one understands the velocity of individual fluid particles along their individual trajectory $\vec{x}(t,\vec{x}_0)$, i.e. $\vec{x}(t_0,\vec{x}_0)=\vec{x}_0$.
It is related to the Eulerian velocity by 
\begin{equation}
 v_L(\vec{x}_0,t)\equiv\partial_t \vec{x}(t,\vec{x}_0)=v(\vec{x}(t,\vec{x}_0),t).
\end{equation}
Consequently spatial correlations, due to the mapping at a given time are equal for Eulerian and 
Lagrangian variables, yet temporal decorrelation are found to deviate between the two formulations.
This defect is often explained in terms of the random sweeping approximation (RSA), which we motivate in the following.

\subsubsection{Random Sweeping Approximation}\hspace*{\fill}
Historically the random sweeping approximation is best understood in the context of \possessivecite{Kraichnan1959} direct interaction approximation (DIA).
The DIA itself is a self-consistent analytical perturbation theory for turbulence that in its original formulation predicts an inertial 
range scaling with $k^{-1/2}$, which is incompatible with the observed and predicted Kolmogorov spectrum $k^{-2/3}$ for hydrodynamic turbulence.

These equations were later reformulated by integrating over the evolution of individual particle paths, 
which is known as the Lagrangian history DIA (LHDIA) \cite{Kraichnan1965}.
This particular model is in agreement with the Kolmogorov spectrum $k^{-2/3}$.
Due to the equality of the energy spectra in the two formulations and the fact that the shape of the energy spectrum is related to one of these timescales, the LHDIA shows that energy 
transfer can be described by the Lagrangian timescale.
One of the particular deviations between the Eulerian and Lagrangian DIA is that the Lagrangian DIA is invariant under 
``random Galilean transformations'' (RGT), which are transformation of the type $v(k,t)\to \exp(-ikut)v(k,t)$, where $u$ is the velocity of the 
Galilean transformation.
In general the defect of the DIA in this regard can be understood as a failure of the DIA to properly resum an appearing infrared 
divergence.
Therefore, the RGT is considered as an important invariance in incompressible homogeneous isotropic turbulence.

The RSA can consequently be used to estimate decorrelation timescales.
The impact of the RSA can be understood by studying the advection of small scale fluctuations by a large scale velocity field $U$, which we assume
to be constant in time but random in space and we neglect pressure and turbulent 
back-reaction.
This gives
\begin{equation}
 \partial_t \vec{v}(\vec{k},t)=-i\vec{k}\cdot\vec{U}\vec{v}(\vec{k},t) \label{sweepeq}
\end{equation}
and the solution is
\begin{equation}
 \vec{v}(\vec{k},t)=\exp\left[-i\vec{k}\cdot\vec{U}(t-t_0)\right]\vec{v}(\vec{k},t_0).
\end{equation}
Then, for the correlation function we find, by averaging via a Gaussian distribution over the velocities $U$ with variance $\ev{U^2}$,
\begin{equation}
 \ev{\vec{v}(\vec{k},t)\vec{v}(\vec{k},t+\Delta t)}=\exp\left(-\frac{1}{2}k^2\ev{U^2}\Delta t^2\right)\ev{v^2(\vec{k},t)}.
\end{equation}
In general one uses this to define the Eulerian eddy turnover time as $t_E=(k\sqrt{\ev{U^2}})^{-1}$.
The simplest Ansatz for the velocity is the rms-velocity 
\begin{equation}
 \ev{U^2}=\frac{2}{3}\int E(k) {\rm d} \ln(k)=\ev{v_1^2}=\frac{1}{2}\frac{\Omega_{t}}{\Omega_r},
\end{equation}
where $\ev{v_1^2}$ is the variance of the 1-component (e.g. x-direction) of the isotropic velocity field.

Thus, the resulting unequal time correlation function is
\begin{equation}
 f_{\rm RSA}(k,\Delta t,0)=\exp\left[-\frac{1}{2}\left(\frac{\Delta t}{t_E}\right)^2\right] \label{RSAdec}.
\end{equation}
The above sweeping effect (\ref{sweepeq}), i.e. the fact that advection is dominated by the root mean square velocity, has also been observed in a series of experiments \citeaffixed{Favre1965}{e.g.}.
The difference between Lagrangian and Eulerian UTCs can be understood in the following way.
As already mentioned in the fluid particle picture, the evolution of the trajectory is mostly guided by the local flow.

Since local properties of the flow change much faster than that of large scales, we can apply a Gallilei/ Lorentz transformation with respect to the small scale flow and the large scale velocity,
which is approximately $v_1$, due to isotropy.
The respective reference frame is approximately that of a fixed observer, if we average over all directions.

In general the Gaussian function (\ref{RSAdec}) predicts the behavior of the UTC quite well for small times.
However, it has been found in simulations that the functional form of the UTC can differ strongly w.r.t. the Gaussian model at late times.
In particular, the values of the unequal time correlation can become negative, e.g. the true UTC is more similar to a 
damped oscillatory function \cite{Rubinstein2003}.
Although as we argue later, this might due to the fact that many earlier simulations might not have clearly distinguished between solenoidal and dilatational components.

Besides, non-zero helicities  can modify the UTC.
\citeasnoun{Rubinstein1999} has argued that a nonzero kinetic helicity leads to a reduction of the sweeping velocity.
Especially for magnetic helicity one might anticipate not only a shift in the sweeping velocity but in the overall functional shape, due to the inverse cascade.
Ultimately, we assume that the impact of helicity on the eddy turnover time will be small and hence expect that it will not alter 
the results much.
Moreover, \possessivecite{Kaneda1999} developed a scheme based on Pade-approximations to evaluate a more precise UTC, yet this makes calculations much more 
expensive and hence will not be discussed.
Generally, we are not interested in the exact long-time behavior, as we argue later.

For $\ev{U^2}$ we use the estimate by \citeasnoun{Kaneda1993}, since the previously mentioned $\ev{v_1^2}$ Ansatz is only reasonable for small scales, which is based on first order Gaussian closure approximation and 
Taylor series approximation and hence only valid up to and including $\mathcal{O}(t^2)$, although the model discussed here differs only slightly from the simple rms velocity Ansatz.
Note that these calculations do not fully resolve turbulent back-reaction, e.g. correlators like $\ev{v^4}$ are neglected and only
correlators like $\ev{v^2 U^2}$ are taken into account.
It is given as 
\begin{equation}
 \ev{U^2(k)}=\int_{-\infty}^\infty h\left(\frac{q}{k}\right)E(q) {\rm d}\ln(q),
\end{equation}
where
\begin{equation}
 h(x)=\frac{1}{24}\left(13-8x^2+3x^4\right)+\frac{1}{16x}(1-x^2)^3\ln\left[\frac{1+x}{|1-x|}\right].
\end{equation}
In general we use for the Kolmogorov spectrum the following approximation 
\begin{equation}
 \ev{U^2(K)}=\ev{v_1^2}\left(\frac{1+0.26K}{\sqrt{2.5}+0.26K}\right)^2. \label{KanSweep}
\end{equation}
As should be clear from the above expression the variance has a constant value $2/5\sqrt{\ev{v_1^2}}$ in the $K\to0$ and 
$\sqrt{\ev{v_1^2}}$ in the $K\to\infty$ limit.
 \begin{figure}[htbp]
 \centering
 \begin{minipage}{0.45\textwidth}
 \includegraphics[scale=0.55]{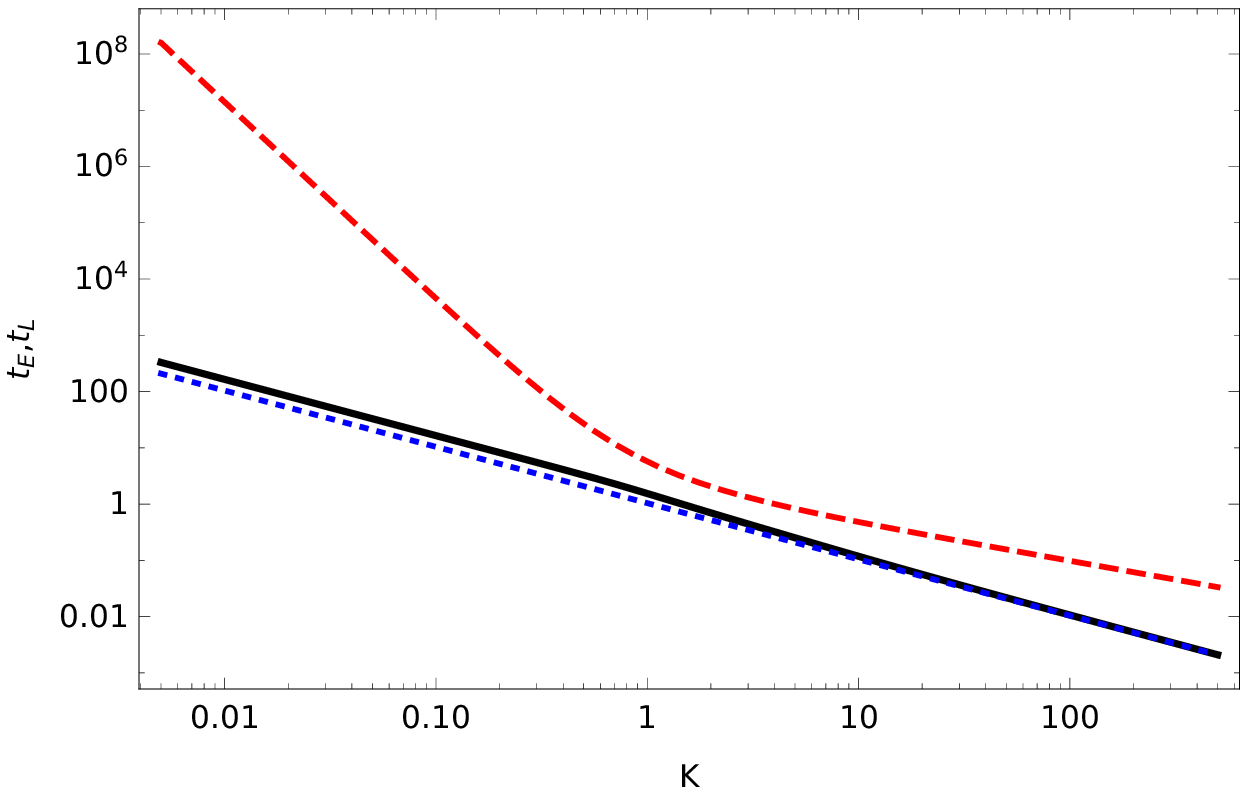}
 \end{minipage}
 \hfill
 \begin{minipage}{0.45\textwidth}
 \includegraphics[scale=0.55]{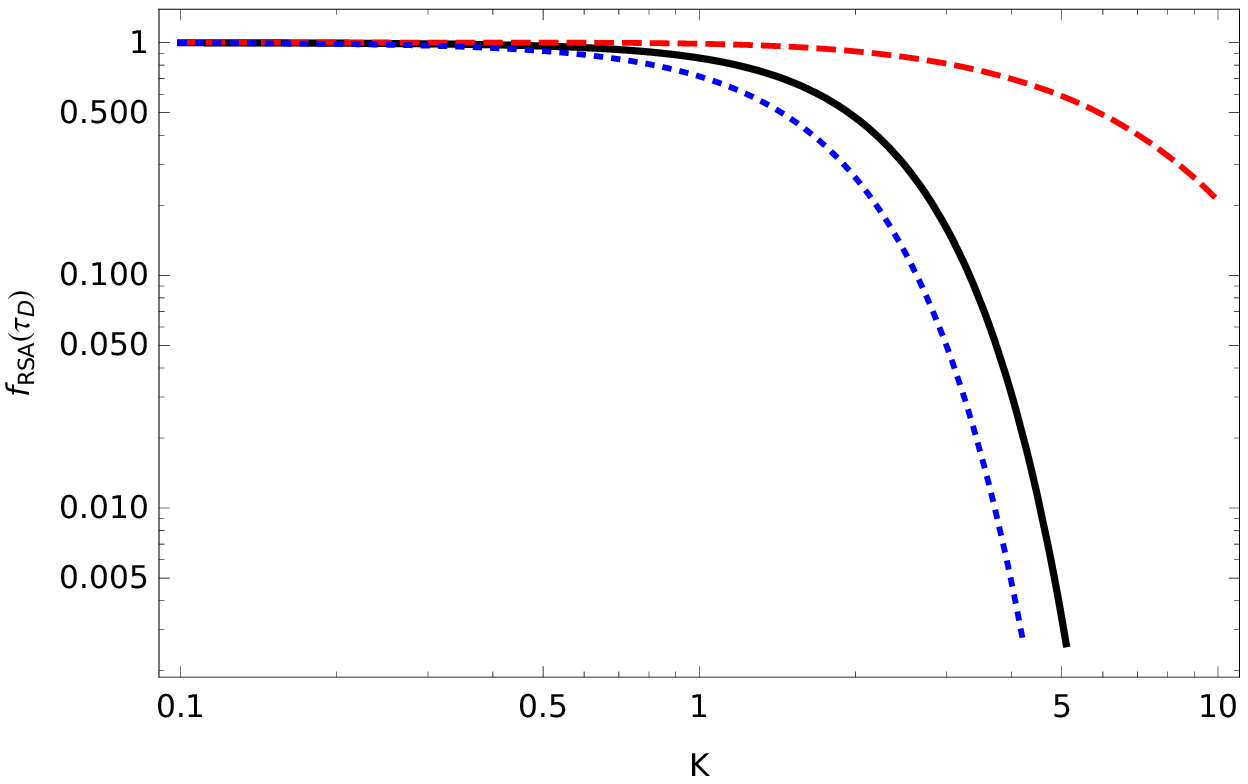}
 \end{minipage}
 \caption{The left panel shows the Lagrangian eddy turnover time (\ref{lagtime}) (red, dashed) and the Eulerian eddy turnover time based on
(\ref{KanSweep}) (black, solid) and based on the $\ev{v_1^2}$ Ansatz (blue, dotted) in arb. units as a function of dimensionless wavenumber $K\equiv kL_I/(2\pi)$.
The right panel shows the Gaussian function (\ref{RSAdec}) $f_{\rm RSA}$, evaluated with the decorrelation timescales shown in the left panel with corresponding line and color styles
at time $\tau=\tau_D$.}
 \label{tempfac}
 \end{figure}
To sum up, the Eulerian eddy turnover time scales as $t_E\propto k^{-1}$, which agrees with simulations and experiments of 
UTCs \citeaffixed{Kaneda1999,Dong2008}{}. 
For Lagrangian variables the eddy turnover time $t_{L}\propto k^{-2/3}$ in the inertial range and $t_L\propto k^{-7/2}$ for a $k^5$ tail
at large length scales. The different timescales are also displayed in figure (\ref{tempfac}) as functions of wavenumber.

Back-reaction caused by terms of order $v^4$ and $b^4$ can be neglected provided that the fluctuations decorrelate on a timescale
$t_E(k)$ small compared to the timescale $t_D$ on which the turbulent amplitude decays. This is the case for $k\gtrsim L^{-1}$,
as can be seen from the right panel in figure (\ref{tempfac}) which shows $\exp\left[-(t_D/t_E(k))^2/2\right]$.
On the other hand, turbulent decay could have a more prominent impact on decorrelation for small $k$, since $t_D\lesssim t_E(k)$ for sufficiently small $k$.
However, for decay that corresponds to a conserved Loitsyansky constant and for $k\lesssim k_I$ the energy density $E(k)$ remains constant for sufficiently long times,
such that the UTC can be approximated sufficiently well by a Gaussian. 
Due to the fact that the spectrum itself follows a $k^5$ power law in the small frequency range, we do not expect that the impact of turbulent decay on the UTC will considerably alter the
gravitational wave emission. Generally MHD simulations\cite{Kaneda1999,LiZhangHe2013} support this qualitative picture.
As is also clear from figure (\ref{tempfac}), using the Lagrangian timescale would considerably change the picture discussed above,
since for $K\gtrsim1$ decorrelation would take significantly longer. 

In 3D homogeneous isotropic turbulence kinetic helicity will amplify the cascade towards smaller length scales, however magnetic helicity drives an inverse cascade and both of these factors will likely increase the 
influence of free decay on the UTC, e.g. we anticipate that the UTC should deviate more quickly from $1$ over a time $\tau_D$ for wavenumbers $k\sim1/L_I$ range.

Further, the inverse eddy turnover rate is effectively $t_E^{-1}\propto k\sqrt{\Omega_t}$. 
Since this can be smaller than the Hubble rate, we introduce a cutoff for contributions from modes with $t_E(k)\gtrsim t_H$.
We argue that eddies with a timescale $t_E(k)\gtrsim t_H$ will not contribute to 
the gravitational wave production, since such a source is effectively constant over a Hubble time time and hence should not contribute to the GW spectrum.
Hence, we further multiply to the UTC the factor $\theta(t_{H}-\chi t_{E}(k))$, where $\chi\sim1$.
The parameter $\chi$ parameterizes the uncertainty in the proposed cut-off criterion.
As we see later on, this factor is only required for maximally helical fields.

Another factor that impacts the decorrelation is a finite viscosity $\nu$ and resistivity $\eta$.
Also, a finite viscosity modifies the RSA, e.g. $\mathcal{O}(t^2)$ contributions are modified \cite{Kaneda1993}.
Yet, as already mentioned we neglect the effect of viscosity altogether.
Consequently, the generalized magnetic field two point correlations are
\begin{equation}
\ev{b_i(\vec{k},\tau')b_j^*(\vec{q},\tau)}=\theta(\tau_{H}-\chi\tau_{E}(k,\tau))\exp\left[-\frac{(\tau'-\tau)^2}{\tau_E^2(k,\tau)}\right]\frac{(2\pi)^6}{4\pi k^2}
\delta(\vec{k}-\vec{q})\left[P_{ij}(\vec{k})E_B(\vec{k},\tau)-i\epsilon_{ijl}\frac{k^l}{k}h_B(\vec{k},\tau)\right], \label{uneqtimecor}
\end{equation}
and the generalized velocity fluctuations are
\begin{equation}
\ev{v_i(\vec{k},\tau')v_j^*(\vec{q},\tau)}=\theta(\tau_{H}-\chi\tau_{E}(k,\tau))\exp\left[-\frac{(\tau'-\tau)^2}{\tau_E^2(k,\tau)}\right]\frac{(2\pi)^6}{4\pi k^2}\delta(\vec{k}-
\vec{q})\left[P_{ij}(\vec{k})E_V(\vec{k},\tau)-i\epsilon_{ijl}\frac{k^l}{k}h_V(\vec{k},\tau)\right], \label{uneqtimecorvel}
\end{equation}
where $a\tau_E=t_E$ is the conformal Eulerian eddy turnover time, $\tau_{H}$ is the conformal Hubble time and recall that $\chi\sim1$ parameterizes the uncertainty in defining a cutoff criterion.
Note, in the above expression we only consider times $\tau'$ and $\tau$ at which the turbulent evolution is described by free decay.

In summary, we have discussed unequal time correlation with a particular focus on the sweeping effect, that is in agreement with observations,
that the rate of decorrelation is the inverse of the Eulerian eddy turnover time $t_E^{-1}\propto k$ and not the often used Lagrangian eddy turnover time, which plays an 
important role in the problem of turbulent energy transport in both formulations.
Although this is only the case as long as the relevant equations are formulated in terms of Eulerian variables.
Overall, unequal time correlations are well approximated by a Gaussian function with rate $t_E^{-1}$.

\subsubsection{Decorrelation of forced incompressible fluctuations}\hspace*{\fill}
\newline
The above analysis only applies to freely decaying turbulence.
For the build-up of the spectrum the build-up of the turbulence also needs to be modelled.
Since we also consider the build-up of the turbulent spectrum, we require an estimate for the decorrelation at those times.
We assume that the bubble collisions that induce the bulk motion can be described as an uncorrelated stochastic process (white noise).
Hence we propose that the build up can be described as random forcing i.e.
\begin{equation}
 \partial_t \vec{v}(\vec{k},t)=\vec{f}(\vec{k},t),
\end{equation}
where $f(k,t)$ is a random force responsible for the development of homogeneous isotropic turbulence.
Since we assume this to behave as white-noise, we have 
\begin{equation}
\ev{f_i(\vec{k},t')f_j(\vec{k},t)}\propto P\delta(t-t'),
\end{equation}
where $P$ is the average amplitude of the force, which we assume to be constant in time.
Via (\ref{enevo}) we find
\begin{equation}
 \frac{\ev{v_i(t,k)v_j(t',k)}}{\ev{v_i(t,k)v_j(t,k)}}=\frac{t'-t_0}{t-t_0}, \label{buildup}
\end{equation}
for $t'<t$ and $t$ and $t'$ are both in the buildup timescale and $t_0$ is the time at which the forcing starts to act.
The above model does not include the sweeping effect and we introduce it at those times, since small scale decorrelation can be much faster than the buildup and hence we assume, for 
a time $t$ during either free decay or forcing and a time $t'$ during forcing, the following UTC
\begin{equation}
  \frac{\ev{v_i(t,k)v_j(t',k)}}{\ev{v_i(t,k)v_j(t,k)}}=\frac{t'-t_0}{t-t_0}\exp\left[-\frac{1}{2}\left(\frac{t-t'}{t_E(t,k)}\right)^2\right]. \label{genbuildup}
\end{equation}
In order to solve the gravitational wave equation, we will choose the condition $h'(t_0)=0$, yet this requires that the source terms vanishes at $t_0$ and hence the above 
extended UTC guarantees that this initial condition is satisfied.
Inconsistent handling of the initial condition can lead to a strongly oscillatory spectrum or negative energies.

\subsubsection{Decorrelation of compressible fluctuations}\hspace*{\fill}
\newline
So far, the previous analysis has only focused on the incompressible UTC.
For compressible flows the above analysis has to be modified.
A compressible flow can generally be separated into two parts, a solenoidal component $\vec{v}_S$, that represents the incompressible part of the flow, and a dilatational component $\vec{v}_D$, 
that represents the purely compressible part of the flow e.g. sound waves.
The dilatational flow itself will be subject to a linear wave propagation, simply due to the propagation of sound waves and the fact that sound oscillations are longitudinal oscillation.
The UTC of dilatational components can be described by an extension of the sweeping model i.e. the swept-wave model \cite{LiZhangHe2013}, which combines the linear wave propagation model \cite{Lee1992} with the 
sweeping approximation.
Then,
\begin{equation}
 \ev{v_D(t,k)v_D(t',k)}=\ev{v_D(t,k)v_D(t,k)}\exp\left[-\frac{1}{2}\left(\frac{t-t'}{t_E(k)}\right)^2\right]\cos\left[c_s k(t-t')\right] \label{sweptwave},
\end{equation}
where the cosine represents the longitudinal fluctuations of the sound waves and the timescale $t_E(k)$ is based on the rms velocity of the solenoidal component of the velocity field.
The above model does not represent solenoidal components, which are still described by the sweeping approximation.

The previously mentioned super-Hubble cutoff is not applicable dilatational fluctuations, since the timescale of variation is the sound crossing timescale which for sub-sound horizon scales fulfills the condition 
$t_{sh}\lesssim t_H$.
Further, if dilatational modes are dominant then nonlinear interaction of dilatational modes might impact decorrelation, as well as shocks even for subsonic flows i.e. $v\gtrsim0.1c_s$ \cite{Kida1990}.
Moreover, \citeasnoun{Rubinstein1999} argued that kinetic helicity will further amplify interactions between sound waves and overall lead to a shift of the sound spectrum to lower frequencies.

For plasmas, the above analysis is incomplete, since in plasmas additional waves are excited e.g. Alfven and magneto-sonic waves.
The Alfven effect is related to the appearance of an Iroshnikov-Kraichnan spectrum in the flow, yet simulations of isotropic and homogeneous turbulence suggest 
that this effect will not be relevant for energy transfer at least in the isotropic case \citeaffixed{Biskamp2000}{e.g.}, although it might be relevant for temporal decorrelation.
Also magnetic fields might amplify a conversion of dilatational to solenoidal modes \citeaffixed{Porter2015}{e.g.}.
Magnetosonic waves are longitudinal waves with phase velocity
\begin{equation}
c_A=\frac{\omega}{k}=\sqrt{\frac{c_s^2+v_A^2}{1+v_A^2}},
\end{equation}
where $v_A$ is a characteristic Alfven velocity and for the case of homogeneous isotropic MHD turbulence we fix $v_A^2=2/3(\Omega_B/\Omega_r)$, where $\Omega_B$ is the magnetic 
field energy density.
Hence, we anticipate that dilatational modes in MHD turbulence follow
\begin{equation}
 \ev{v_D(t,k)v_D(t',k)}=\ev{v_D(t,k)v_D(t,k)}\exp\left[-\frac{1}{2}\left(\frac{t-t'}{t_E(k)}\right)^2\right]\cos\left[c_A k(t-t')\right].
\end{equation}
The above model has to be considered with some caution, since magnetosonic waves are generally associated with some mean field $\vec{B}_0$ and the assumption that
the relevant Alfven velocity corresponds to that of purely stochastic fields is not as obvious as for velocity fluctuations.
Because magnetic fields are not invariant under "Alfven Galilean transformations" i.e. a Galilean transformation of magnetic fluctuations with respect to an Alfven wave. 
Indeed, an Iroshnikov-Kraichnan spectrum is typically observed in simulations for anisotropic turbulence with a nonzero mean magnetic field for the perpendicular 
components \cite{Muller2005}, confirming the assertion that unlike for hydrodynamic turbulence, the impact of a non-zero mean value on the spectrum is considerable i.e. there is 
no extension of the simplistic RGT picture to magnetic field fluctuations.
A related problem is the presence of magnetic sweeping, i.e. a contribution of magnetic fields to $t_E(k)$.
Here we presume that this is the case and assume that the kinetic sweeping effect is extended by a magnetic sweeping effect with velocity $b_1$ i.e. the sweeping is driven 
by the dominant contribution and the relevant value for sweeping is $\Omega_t=\max(\Omega_B,f_S\Omega_V)$.
Note that this assertion is less certain and generally will require further investigation.
In general decorrelation properties of hydrodynamic, isotropic, homogeneous and incompressible turbulence are quite well understood, but for homogeneous isotropic compressible MHD turbulence 
there are still some uncertainties.

In summary in this section we covered some basics of MHD turbulence in the early universe, the basic spectra, turbulent evolution and we in particular focused on turbulent decorrelation and the relevant
timescales involved. In particular we have discussed the difference between the Lagrangian and Eulerian eddy turnover time and also how decorrelation can differ for solenoidal and dilatational modes, 
since the decorrelation is primarily due to the solenoidal modes and hence dilatational modes remain correlated longer.

\section{Gravitational wave production by MHD turbulence}
In this section, the general solution to (\ref{gw-eq2}) or (\ref{gw-eq3}) is discussed in the case of a radiation 
dominated universe by taking into account the previously defined source terms and UTC.

\subsection{Gravitational wave energy density}
In the radiation dominated phase the scale factor, expressed in conformal time, is ${\rm d}\tau={\rm d} t/a $, where $H_0$ is the Hubble parameter at $a_0=a(\tau_0)$ at time 
$\tau_0$ i.e. today.
Thus, (\ref{gw-eq3}) reduces to
\begin{equation}
\left(\partial_\tau^2+\frac{2}{\tau}\partial_\tau+k^2\right)h_{ij}^T(\vec{k},\tau)=16\pi G_{\rm N}\pi_{ij}^T(\vec{k},\tau) 
\end{equation}
with the initial conditions $h_{ij}(\tau_0,\vec{k})=\partial_\tau h_{ij}(\tau_0,\vec{k})=0$.
The solution via the relevant Green's function is
\begin{equation}
h_{ij}(k,\tau)=16\pi G_{\rm N}\left[A_{ij}(k,\tau)\frac{\sin(k\tau)}{k\tau}-B_{ij}(k,\tau)\frac{\cos(k,\tau)}{k\tau}\right],
\end{equation}
where
\begin{equation}
A_{ij}(k,\tau)=\int_{\tau_0}^\tau\pi_{ij}^T(k,\tau')\cos(k\tau')\tau'{\rm d}\tau'
\end{equation}
and
\begin{equation}
B_{ij}(k,\tau)=\int_{\tau_0}^\tau\pi_{ij}^T(k,\tau')\sin(k\tau')\tau'{\rm d}\tau'.
\end{equation}
Then the change of the strain is
\begin{equation}
 \partial_\tau h_{ij}(k,\tau)=\frac{16\pi G_{\rm N}}{k\tau^2}\left[A_{ij}(k,\tau)\left(k\tau\cos(k\tau)-\sin(k\tau)\right)+B_{ij}(k,\tau)
 \left(k\tau\sin(k\tau)+\cos(k\tau)\right)\right]. \label{changestrain}
\end{equation}

Next, the comoving gravitational wave energy density seeded by magnetic fields in a radiation dominated universe is
\begin{align}
 \rho_G(\vec{x},\tau)=&8\pi G_{\rm N}H_0^2\Omega_r f_g\int{\rm d}^3\vec{k}\int\frac{{\rm d}^3\vec{k}'}{(2\pi)^3}
 e^{-i(\vec{k}-\vec{k}')\cdot\vec{x}}\int_{\tau_0}^\tau {\rm d}\tau'\
 \int_{\tau_0}^\tau {\rm d}\tau'' \tau'\tau''F_{\rm RD}(k,k',\tau,\tau',\tau'') \\
 &\times \sum_{ij}\ev{\pi_{ij}^T(\vec{k},\tau') \pi_{ij}^{T*}(\vec{k}',\tau')},\nonumber
 \end{align}
 where $f_g=(g_0/g_*)^(1/3)$ represents the ratio of particle degrees of freedom with $g_0=3.36$ and $g_*$ the degree of freedom at temperature $T_*$ e.g. the phase transition and 
$F_{\rm RD}(k,k',\tau,\tau',\tau'')\approx\cos(k(\tau-\tau'))\cos(k'(\tau-\tau''))$ for details see appendix A.

Focusing on the term $\ev{\pi_{ij}^T(\vec{k},\tau') \pi_{ij}^{T*}(\vec{k}',\tau')}$, we see that we require the knowledge of 
quartic correlators e.g. $\ev{v_iv_jv_kv_l}$.
Therefore we assume that the fluctuations can be approximated as normal distributed.
Consequently we can now evaluate the correlations using Wick's theorem
\begin{equation}
\ev{A_a B_b C_c D_d}=\ev{A_a B_b}\ev{C_c D_d}
+\ev{A_a C_c}\ev{B_b D_d}+\ev{A_a D_d}\ev{C_c B_b}, \label{Wickt}
\end{equation}
where $A$, $B$, $C$ and $D$ are normal distributed vector or pseudo-vector fields.
For details on this, see appendix A.
 
Besides, we neglect correlators of the form $\ev{vB}$, as these are in the context of ideal MHD conserved quantities 
(at least $\ev{\vec{v}\cdot\vec{B}}$ is conserved).
Hence if we assume that these cross correlations are 0 initially, they will remain 0.
Additionally we define several quantities 
\begin{align}
 E_t^2(q,p,\tau)&=E_S(q,\tau)E_S(p,\tau)+E_B(q,\tau)E_B(p,\tau) \label{Et}\\
 H_t^2(q,p,\tau)&=h_B(q,\tau)h_B(p,\tau)+h_V(q,\tau)h_V(p,\tau) \\
 S^\pm(k,q,p)&=1+\frac{1\pm3}{2}\left[(\hat{k}\cdot\hat{q})^2+(\hat{k}\cdot\hat{p})^2\right]+(\hat{q}\cdot\hat{k})^2(\hat{k}\cdot\hat{p})^2 \label{spmcoff}\\
 D(k,q,p)&=(\hat{k}\cdot\hat{p})^2\left(1-(\hat{q}\cdot\hat{k})^2\right)\label{dcoff},
\end{align}
where the terms $S^{\pm}(k,q,p)$ and $D(k,q,p)$ result from the product of the correlation function with the projectors and their derivation is shown in appendix A.
Next, we apply Wick's theorem, evaluate the product of the projectors and apply (\ref{uneqtimecor}) and (\ref{uneqtimecorvel}) and bring the equations in a time ordered form, as shown in appendix A,
to arrive at 
 \begin{align}
 \rho_G(\vec{x},\tau)\approx& \frac{8\pi}{(4\pi)^2} G_{\rm N}H_0^{-2}\Omega_r f_g(\rho+p)^2\int{\rm d}^3\vec{k}\int{\rm d}^3\vec{q}
 \int_{\tau_0}^\tau {\rm d}\tau'\ \int_{\tau_0}^{\tau'} {\rm d}\tau''\
  \frac{1}{q^3 p^3\tau'\tau''} \cos(k(\tau'-\tau'')) \nonumber \\
 &\times f_{\rm RSA}(\tau',\tau'',q)f_{\rm RSA}(\tau',\tau'',p)\Bigl[E_t^2(q,p,\tau')S^+(k,q,p)+4H_t^2(q,p,\tau')(\hat{k}\cdot\hat{q})(\hat{k}\cdot\hat{p})+\nonumber\\
& S^-(k,q,p)\Bigl(4E_D(q,\tau')E_D(p,\tau')\cos\left[qc_s(\tau'-\tau'')\right]\cos\left[pc_s(\tau'-\tau'')\right]\Bigr)+\nonumber\\
 &6D(k,p,q)E_S(q,\tau')E_D(p,\tau')\cos\left[pc_s(\tau'-\tau'')\right]+\nonumber\\
 & 6D(k,q,p)E_S(p,\tau')E_D(q,\tau')\cos\left[qc_s(\tau'-\tau'')\right]\Bigr].  
  \label{gwenord}
 \end{align}

Moreover, we are interested in the gravitational wave power spectrum, which we find by neglecting the integration over $k$ as seen in
(\ref{powspec}).
Therefore the GW power spectrum is found by setting $\int{\rm d}^3k\to 4\pi k^3$ and dividing by the critical energy density in
(\ref{powspec}).
Ultimately, we get
 \begin{align}
 \Omega_{GW}(\vec{k},\tau)\approx&\left(\frac{g_0}{g_*}\right)^{1/3}\frac{16\Omega_r}{3}\frac{k^3}{4\pi}\int{\rm d}^3\vec{q}
 \int_{\tau_0}^\tau {\rm d}\tau'\ \int_{\tau_0}^{\tau'} {\rm d}\tau''\
  \frac{1}{p^3 q^3 \tau''\tau'} \cos(k(\tau'-\tau'')) \nonumber \\
 &\times f_{\rm RSA}(\tau',\tau'',q)f_{\rm RSA}(\tau',\tau'',p)\Bigl[E_t^2(q,p,\tau')S^+(k,q,p)+4H_t^2(q,p,\tau')(\hat{k}\cdot\hat{q})(\hat{k}\cdot\hat{p})+\nonumber\\
& S^-(k,q,p)\Bigl(4E_D(q,\tau')E_D(p,\tau')\cos\left[qc_s(\tau'-\tau'')\right]\cos\left[pc_s(\tau'-\tau'')\right]\Bigr)+\nonumber\\
 &6D(k,p,q)E_S(q,\tau')E_D(p,\tau')\cos\left[pc_s(\tau'-\tau'')\right]+\nonumber\\
 & 6D(k,q,p)E_S(p,\tau')E_D(q,\tau')\cos\left[qc_s(\tau'-\tau'')\right]\Bigr]. \label{gwpowspec}
 \end{align}
In the following we first analyze these equations assuming the initial conditions and scaling properties discussed so far.

 \subsection{General scaling properties of the MHD-GW-equation}
 
 A simplistic picture of the overall scaling properties of the derived gravitational wave spectrum from MHD turbulence can be
 estimated as follows.
 In this analysis we distinguish two cases $k\ll 2\pi/L_I=k_I$ and $k\gg k_I$. 
 Also we neglect helicity.
 First, we start by assuming $\tau_E(q)\ll \tau_H$ for the relevant modes $q$.
 Then we estimate the time integration over $\tau''$ in equation (\ref{gwpowspec}) by restricting the integration range
 to the interval $[\tau'-\tau_E(\tau',q),\tau']$ in the case that $\tau'-\tau_E(\tau',q)>\tau_0$, which we assume to be the case for the most 
 relevant time and length-scales (turbulence formed maximally around initial integral scale).
 Additionally we constrain the total integration time $\tau$ by a few turbulent decay times $\lambda\tau_D$, where $\lambda$ is a 
 parameter of $\mathcal{O}(1)$.
 Further, the spatial integration can be written as
 \begin{equation}
\int{\rm d}^3\vec{q} f(q,p,k)= 2\pi\int_0^\infty {\rm d} q \int_{|q-k|}^{q+k}{\rm d} p \frac{pq}{k} f(q,p,k),  
 \end{equation}
where $p=\sqrt{|\vec{q}-\vec{k}|}$ and $f(q,p,k)$ depends only on the absolute of the three wavenumbers.
Since the biggest contribution comes from modes $q\sim k_I$ and $k\ll k_I$ we can simply perform the $p$ integration, by setting 
$p\to q$.
Then, the width of the $p$ integral is a factor $2k$ and the factor coming from the product of projectors is approximately
\begin{equation}
 \frac{1}{2}\int_{-1}^{1}{\rm d}\cos(\theta)\left[1+2(\hat{k}\cdot\hat{q})^2+2(\hat{k}\cdot\hat{p})^2+(\hat{q}\cdot\hat{k})^2(\hat{k}\cdot\hat{p})^2\right]\to \frac{38}{15}.
\end{equation}
Here, we use the simplified $\tau_E^{-1}=k\sqrt{\ev{v_1^2}}\equiv ku$ model, as it provides a simple and reasonable approximation to the wavenumber dependence of the more advanced model based on (\ref{KanSweep}).
Approximating the UTC function by its Taylor series up to $\mathcal{O}((\tau'-\tau'')^3)$ and integrating over the reduced integration range while neglecting the 
variation of the $1/\tau''$ factor, due to near constancy, we find
\begin{equation}
 \Omega_{GW}(\vec{k},\tau)\propto k^3\int_0^\infty \frac{{\rm d} q}{q^5}
 \int_{\tau_0}^{\tau_0+\lambda\tau_D} {\rm d}\tau'\frac{1}{\tau'^2} E(q,\tau')E(q,\tau')\left(-u\cos(u^{-1})+\frac{1+2u^2}{2}\sin(u^{-1}))\right). \label{compscal}
\end{equation}
Already, it is clear that the GW spectrum scales as $k^3$ for $k\ll k_I$. 
This is due to the fact that because of causality the source term cannot depend on super-Hubble modes.
And normalizing the appearing scales as $k\to K$ and $q\to Q=qL_I/(2\pi)$ gives an additional factor $L_I$.
The last spatial integration will be more difficult due to terms like $\cos(u^{-1})$, since $u\propto\sqrt{\Omega_t(\tau')}$.

If we neglect the relevance of the cosine (which is questionable) and the UTC in the reduced $\tau'$ integration, the $\tau'$ integration simply gives a factor $\tau_E$.
Moreover, $\tau_E(\tau',q)\sim 1/(q\sqrt{\Omega_t(\tau')})$ and $E(q,\tau')^2\propto (\Omega_t(\tau'))^2$.
Combining the above considerations, we approximate the time integration as
\begin{equation}
 \int_{\tau_0}^{\tau_0+\lambda\tau_D} \frac{{\rm d}\tau'}{\tau'^2} \Omega_{t,*}^{3/2}(\tau')L_I(\tau').
\end{equation}
Now, we take into account the impact of the self-similar decay i.e. (\ref{enevo}) and (\ref{levo}) which gives
\begin{equation}
 \int_{\tau_0}^{\tau_0+\lambda\tau_D} \frac{{\rm d}\tau'}{\tau'^2} \Omega_{t,*}^{3/2}L_*\left(\frac{\tau'-\tau_0}{\tau_D}\right)^{\gamma_1-3/2\gamma_2}.
\end{equation}
In the integration, we neglect the evolution of the $1/\tau'^2$ contribution, since we assume $\tau_D\ll \tau_0$ and this effectively gives a factor $\tau_D$.
Hence for the scaling of the energy, we expect
\begin{equation}
\Omega_{GW}(\vec{k},\tau)\propto \Omega_{t,*} L_*^2 k^3\ \ {\rm for}\ \ k\ll k_I.
\end{equation}
In our numerical calculations we observe a slightly different scaling, that can be approximated as a power law due to only a slight variation of the power law index i.e. 
\begin{equation} 
\Omega_{GW}(\vec{k},\tau)\propto \Omega_{t,*}^{3/2} L_*^2 k^3\ \ {\rm for}\ \ k\ll k_I.
\end{equation}
This different scaling behavior is not unexpected, due to the relevance of the oscillating terms in (\ref{compscal}) that imply a nontrivial dependence on $\Omega_{t,*}$.

If neither free decay nor a nontrivial UTC are impacting the evolution i.e. for sound waves, then we anticipate a different scaling of $L_*\Omega_{t,*}^2$, since the integration timescale in the $\tau''$ 
no longer depends on $q$ and since $k$ is small the integration over the cosine is trivial, without any additional factor $\Omega_{t,*}$ and $L_*$ appearing.
 
Now, we focus on scales  $k\gg k_I$.
Unlike before, we only evaluate the dependence on $k$ here, but note that the $L_*$ dependence is the same.
A key difference is that now the scales of interest are $q\sim k_I$ and $p\sim k$ and due to symmetry also $p\sim k_I$ and $q\sim k$.
For simplicity we only look at the first scales of interest and assume that the contribution, due to symmetry from the other set of scales, will be similar.
Then, performing the $p-$integration evaluates the integrand at $p\sim k$ with width $2q$ and the timescale of interest for the UTC is now evaluated at scale $k$.
Summarizing, we find for the GW spectrum at large scales
\begin{equation}
 \Omega_{GW}(\vec{k},\tau)\propto \int_0^\infty \frac{{\rm d} q}{q}
 \int_{\tau_0}^{\omega\tau_E} \frac{{\rm d}\tau'}{\tau'^2} \tau_E(\tau',k) E(q,\tau')E(k,\tau')\ \ {\rm for}\ \ k\gg k_I.
\end{equation}
Since, $\tau_E(\tau',k)\propto k^{-1}$, the $k-$dependence of the spectrum is simply given by the turbulent energy spectrum $E(k)/k$.

If the UTC is simply 1 over the relevant integration range (i.e. $\tau_H$), the scaling will again differ.
The double integral over $\cos(k(\tau'-\tau''))$ implies a factor $k^{-2}$, hence for such a case one expects that the GW spectrum follows a $E(k)/k^2$ law.
This is e.g. to be expected for a strongly dilatational spectrum, for which the impact of Gaussian decorrelation is negligible.

For a Kolmogorov spectrum $k^{-2/3}$, we anticipate that the GW spectrum follows a ($k^{-5/3}$) spectrum.
Consequently, if $E(k)$ has a Iroshnikov-Kraichnan spectrum $k^{-1/2}$ or a $k^{-1}$ (i.e. sound waves) spectrum, then the GW spectrum will have a $k^{-3/2}$ or $k^{-2}$ spectrum up to the dissipation scale.
In the direct numerical evaluation of the integral, we find that for small values of $\Omega_{t}$ the spectrum develops a $k^{-8/3}$ scaling, which might be indicative of the forcing 
factor due to the term $(\tau''-\tau_0)/(\tau'-\tau_0)$ which at large scales will be simply integrated up to $\tau'$ and will give at most a $\mathcal{O}(1)$ factor but at small scales it 
implies a factor $\tau_E^2(\tau',k)$ and also a more complicated scaling with $\Omega_t$. 

\section{Turbulence from Cosmological Phase Transitions}
Here, we only briefly mention some key aspects of cosmological phase transitions (CPT) of relevance to our analysis.
For a more in depth review, we refer the reader to the recent review by \citeasnoun{Caprini2016}.
A first order phase transition is described by two phases with strongly differing thermodynamic properties.
The new phase generally appears in the form of bubbles that are expanding until the entire fluid is in the new phase.
Then, the expanding bubbles with bubble wall velocity $v_w$ will collide with other bubbles.
One typically distinguishes the bubble expansions into two general classes depending on $v_w$.
Deflagration bubbles are subsonic $v_w\ll c_s$, while detonation bubbles are supersonic $v_w\gg c_s$ \cite{Kaminonkowski1994}.
For $v_{w}\sim c_s$ a clear distinction is not possible and this is typically refereed to as a hybrid scenario \cite{EspKonSer2010}. 
In detonations, the shock front of the bubble is highly compressed and thin and behind the shock wave turbulence might develop.
Also, bubble wall instabilities can source magnetic fields \citeaffixed{Sigl1997}{e.g.}.
These seed fields can then be amplified by the turbulent flow.

Further, one important parameter of the phase transition is the bubble nucleation rate $\beta$. 
The timescale $\beta^{-1}$ is roughly equal to the duration of the phase transition,
i.e. the time after which the appearance of new bubbles ceases and the fluid enters the new phase.
The rate $\beta$ also allows us to estimate the typical size of bubbles at collisions $R_*\sim v_w\beta^{-1}$.
A subscript $*$ generally indicates variables at the phase transition.
The integral scale for the turbulent motion can be estimated as $L_I\sim 2R_*$, since that is the scale where typically most of the turbulent motion will be generated. 

Furthermore, another important parameter is $\alpha=\epsilon_l/\rho_r$ i.e. the ratio of the of latent heat to the background energy density prior to the phase transition.
Consequently, $\alpha$ parameterizes the amount of energy that is released due to the transition to the new equilibrium state (vacuum) and hence $\alpha$ is a measure for 
the strength of a phase transition.
The parameter $\alpha$ implies a further distinction of two scenarios with $\alpha\lesssim1$ and $\alpha\gg1$.
For $\alpha\lesssim1$ the evolution of the background is overall unaffected and the phase transition is called thermal.
But for $\alpha\gg1$ the evolution of the background is strongly driven by the scalar field that is driving the evolution to the new ground state, also referred to as a vacuum phase transition.
Hence, the background evolution is exponential until the new ground state is reached and the scalar field decays during a phase of reheating.
In general only a fraction $\kappa_v$ of the energy in a phase transition is transformed into bulk motion.
In cases with $\alpha\gg1$, one expects $\kappa_v\sim0$. 
For thermal phase transitions $\alpha\lesssim1$, $\kappa_v$ can be estimated for two asymptotic cases \cite{EspKonSer2010}
\begin{equation}
\kappa_v=\begin{cases} \alpha(0.73+0.083\sqrt{\alpha}+\alpha)^{-1} \ \ \quad\qquad v_w\sim1\\
v_w^{6/5}6.9\alpha(1.36-0.037\sqrt{\alpha}+\alpha)^{-1} \quad v_w\lesssim0.1.
\end{cases}
\end{equation}
Moreover, $\kappa_v$ attains a maximum in the hybrid case i.e. $v_w\sim c_s$ and does not peak at $v_w=1$, so for $v_w\sim c_s$, $\kappa_v$ might e.g. be larger by a factor 4 in comparison to the 
value of $\kappa_v$ for $v_w=1$.
Recently, \citeasnoun{CaiWang2018} showed that the efficiency parameters decrease if the expansion of the background is taken into account for slow phase transitions $\beta\sim\mathcal{O}(1)H_*$.

Here we are primarily interested in thermal phase transitions with $\alpha\lesssim1$.
\possessivecite{Hindmarsh2017} showed that for a weak phase transition $\alpha\lesssim 0.1$ solenoidal motion will not be sourced immediately, yet simulations for medium strength phase transitions indicate
that rotational modes become important for strong phase transition, $\alpha\gtrsim0.1$.
Otherwise dilatational modes (sound waves) will become dominant and drive the initial evolution of the turbulent flow.
In general the collisions of bubbles itself sources gravitational waves as well \citeaffixed{HuberKonstandin2008}{e.g.}, but for thermal phase transition, the GW signal from compressible MHD turbulence is 
typically dominant \cite{Caprini2016}.
Furthermore, even if the initial turbulence is dominated by compressional modes, one can expect that for a subsonic flow solenoidal modes will develop due to the interaction of dilatational modes with solenoidal 
modes or magnetic fields or through a nonzero baroclinity $\nabla\rho\times\nabla p\neq0$.

As already mentioned in the introduction, there are at least two particular transitions of interest with respect to the early universe, that is the QCDPT and EWPT.
Both of these transitions might be first order phase transitions.
For the EWPT beyond the standard model physics (BSM) is required \citeaffixed{Laine1999}{e.g.}, whereas the QCD phase transition might be a first order phase transition 
e.g. depeding on the lepton asymmetry \cite{SchwarzStuke2009}.
Some BSM scenarios involve additional scalar fields like the Higgs portal scenario \cite{EspKonRiv2012}, variations of fermionic Yukawa couplings \cite{BalesKonSer2016} and supersymmetric models 
\citeaffixed{Apreda2002}{e.g.}.
From now on we primarily assume $v_w=0.9$, unless specified otherwise, and for $\kappa_v$ we take the value that corresponds to $v_w=1$ for all the cases we study here and focus as an example only on EWPT scenarios 
which are of interest for LISA.

As discussed by \citeasnoun{Caprini2016} some dark sector extensions to the standard model that lead to a first order PT might have phase transition parameters in the range $\alpha\sim0.1\to1$ and $\beta\sim10\to100H_*$ 
and the dark sector PT might be in a temperature range of $100$TeV to $10$MeV.
Also SUSY extensions, in particular singlet extensions to the MSSM, can lead to EWPTs with $\alpha\sim0.1$ and $\beta\sim5\to100H_*$.
The same goes for additional scalar extensions to the standard model, like the Higgs portal scenario \citeaffixed{EspKon2008,Caprini2016}{e.g.}.
Although, it is generally not fully clear which scenarios lead to a thermal or to a nonthermal runaway phase transition ($\alpha\gg1$), for which turbulence will likely not be relevant at all.
Hence, parameter ranges with $\alpha\gtrsim0.1$ and $\beta\sim5\to1000H_*$ are generally of interest and we will discuss them here.

\section{Results}

\begin{figure}[htbp]
 \centering
\includegraphics[scale=1.0]{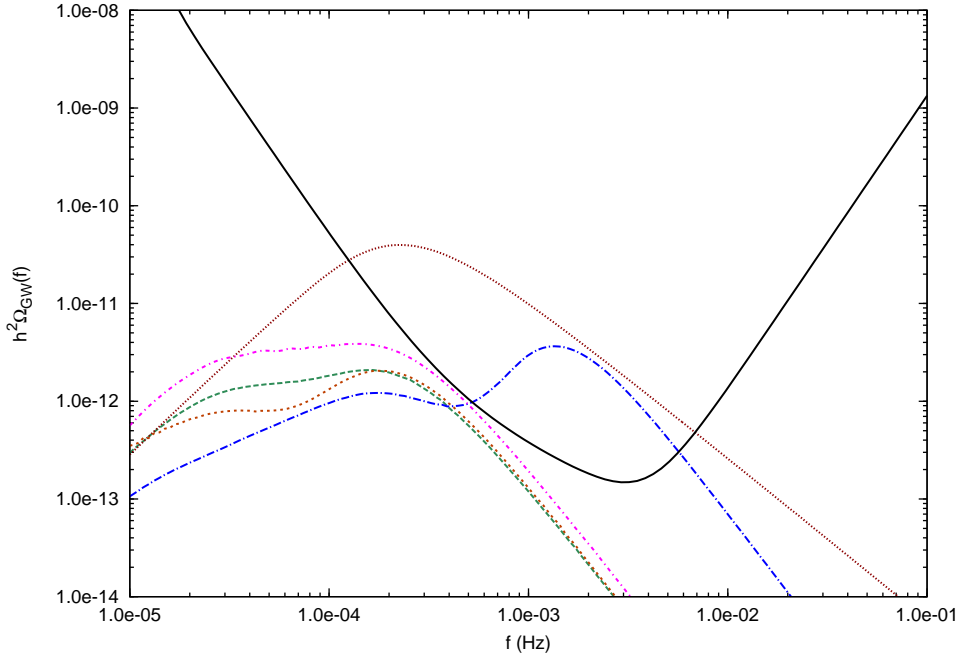}
\caption{The gravitational wave spectrum for the Higgsportal scenario with $\alpha=0.17$, $\beta/H_*=12.5$, $T_*\approx60$GeV. The lines denote the LISA sensitivity curve (black, solid), the so-far used top hat UTC model 
(dark-red, dotted), the Lagrangian UTC (blue, dash-dotted) and the Eulerian UTC model (green, dashed). Further we also consider contributions from modes with timescale $\tau_E(k)>\tau_H$
$t(k)\gtrsim t_H$ ($\chi=0$).
At observable frequencies our calculations based on the sweeping model thus predict an amplitude smaller by roughly a factor 10 compared to the top hat and Lagrangian UTC models.
This is mostly due to the shorter correlation timescales in the Eulerian formulation.
The two other lines indicate two particular enhancements, the magenta line (dot-dashed) shows the spectrum for the case $\tau_{\rm b}=\beta^{-1}$, whereas the buildup times in the other cases are based on the Eddy turnover time.
Lastly, the thick dotted dark-orange line shows the case for maximal magnetic helicity with ($\chi=1$).}
\label{higgspp}
\end{figure}

Now, we present and analyze our calculations for the GW spectrum.
The peak frequency of the gravitational wave spectrum is given by \cite{Caprini2018}
\begin{equation}
 f_{GW}=2.6\cdot 10^{-8} x_{k}\left[\frac{g_*(T_p)}{100}\right]^{\frac{1}{6}}\left(\frac{T_*}{{\rm GeV}}\right){\rm Hz},
\end{equation}
where $x_k=2\pi k_I/H_*$ with $k_I/H_*$ representing the normalized integral scale of the gravitational wave spectrum at the phase transition characterized by $a_*$ and temperature $T_*$.
As before, the subscript $*$ refers to the variables at the phase transition.
First, we discuss one particular scenario, in which a significant fraction of the bubble energy might transform into bulk motions of the fluid, i.e. the Higgs portal scenario \citeaffixed{EspKon2008,Caprini2016}{e.g.}.
We show this scenario in figure (\ref{higgspp}) with $\alpha=0.17$, $\beta/H_*=12.5$, $T_*\approx60$GeV and $v_w=1$ (slightly different values e.g. $v_w=0.9$ only marginally change the result).
For the LISA sensitivity curve, we use the one used in \citeasnoun{Caprini2018} and based on \cite{ThraneRomano2013,Moore2015,LISA2017}.
In figure (\ref{higgspp}) one clearly sees that the Lagrangian model produces significantly more gravitational waves and at higher frequencies than the sweeping model, 
but most of the significant contributions are from modes with $\tau_L>\tau_H$.
Besides, a non-zero $\chi$ (cut-off of modes with $\chi\tau_E(k)<\tau_H$) produces a slight modulation of the spectrum at small frequencies, i.e. a small suppression at scales larger than the gravitational wave integral scale.
Further, it is noticeable that the top-hat model significantly overestimates the gravitational wave energy in comparison to the Eulerian UTC model, but also 
in comparison to the Lagrangian model.
Additionally, the peak frequency is shifted to smaller frequencies in comparison to both the Lagrangian and top hat model.
Moreover, we do not evaluate any negative energies for the Lagrangian decorrelation model, as discussed in \cite{CapDurSer2009}, this is primarily due to our enforcing of the $h'(\tau_0)$ remaining 0 with respect 
to the decorrelation function, otherwise we also found negative energies in the spectrum.
Again,the Lagrangian model is an equally valid description of turbulence, since it should produce the same spectrum as the Eulerian model does, however it cannot be directly applied to the equations developed within the
cosmological rest frame.
Overall, it is unlikely that reasonable deviations from the Gaussian function will significantly alter the shape of the spectrum.
We do not consider Lagrangian decorrelation any further.
For magnetic helcitiy, the overall cascade properties are different in the sense that the decay is more slower and $\Omega_t(\tau)/L_I(\tau)={\rm const}$.
In the Higgs-portal scenario, magnetic helicity would primarily leads to a decrease at frequencies smaller than the peak frequency and a minor increase at larger frequencies.
One particular important factor is the build-up timescale $\tau_{\rm b}$ as in this particular case, the choice $\tau_{\rm b}=\beta^{-1}$ leads to an amplification of the spectrum by around a factor of 2 in
comparison to the case $\tau_{\rm b}=\tau_D$.

\begin{figure}[htbp]
 \centering
 \begin{minipage}{0.45\textwidth}
 \includegraphics[scale=0.55]{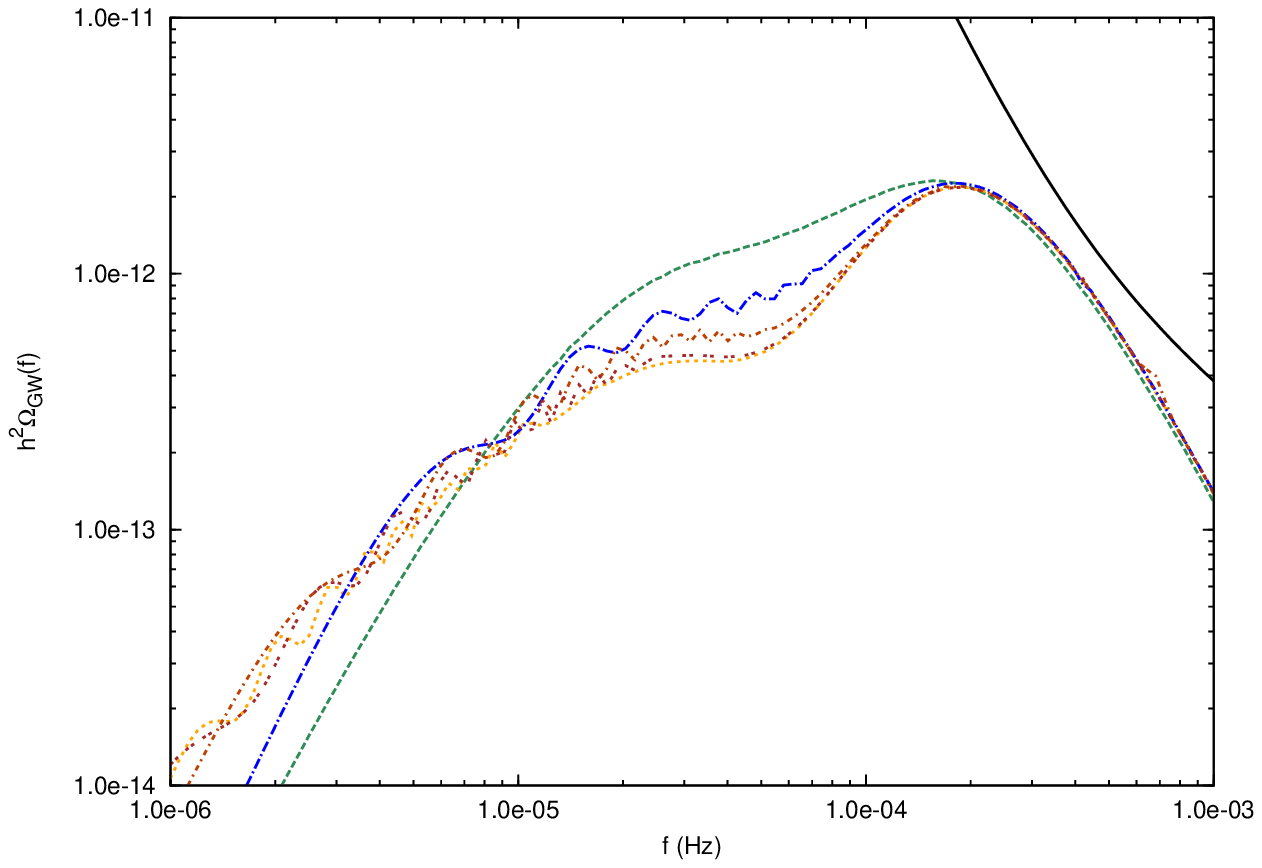}
 \end{minipage}
 \hfill
 \begin{minipage}{0.45\textwidth}
 \includegraphics[scale=0.55]{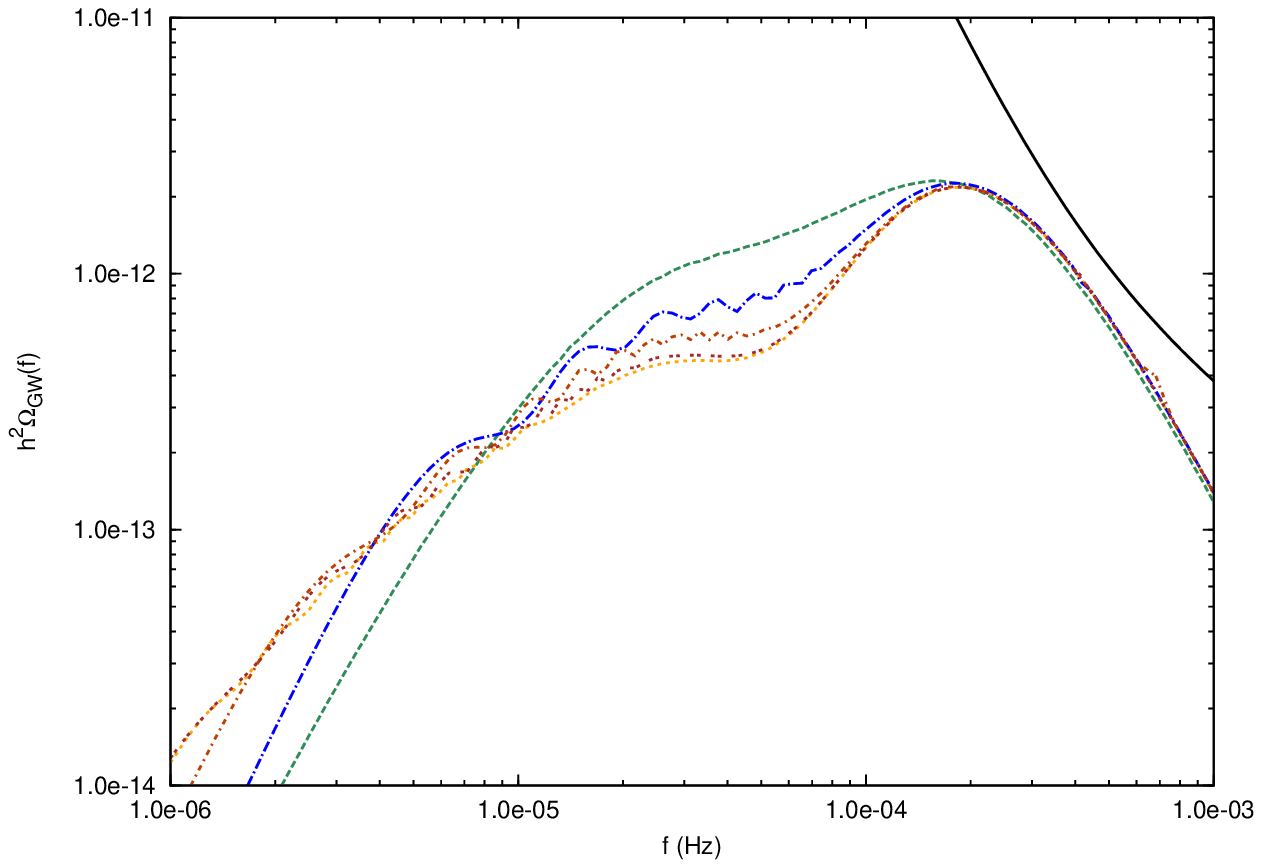}
 \end{minipage}
\caption{The gravitational wave spectrum for the Higgsportal scenario with $\alpha=0.17$, $\beta/H_*=12.5$, $T_*\approx60$ and maximal magnetic helicity for $\chi=0$ (left panel) and $\chi=2$ (right panel). 
The lines denote the LISA sensitivity curve (black, solid), the Eulerian UTC without helicity (green, dashed) and different integration times: $\tau_{max}=3\tau_*$ (blue dot-dashed), 
$\tau_{max}=5\tau_*$ (dark-orange, dot-dashed), $\tau_{max}=10\tau_*$ (brown, double dot),$\tau_{max}=20\tau_*$ (orange-yellow, dotted). The oscillations that appear are generally caused by modes with $\tau_E(k)>\tau_H$ due to the cosine appearing 
in (\ref{gwpowspec}) and also a consequence of the structure function of helical terms $\propto \left(\hat{k}\cdot\hat{p}\right)\left(\hat{k}\cdot\hat{q}\right)$.}
\label{heliplo}
\end{figure}

Now, we discuss the problem of inverse cascades in more detail.
As mentioned in chapter 3, the Eulerian rate of decorrelation is expected to differ for a nonzero helicity and in practice using the Eulerian deccorelation rate discussed here, this will lead to some problems especially when $\tau_E\sim\tau_H$.
In particular the energy growth rate can be negative, which is unphysical since no backreaction terms are taken into account e.g. conversion of gravitational wave energy into turbulent energy.
For $\tau_E\ll\tau_H$ this is less of an issue, yet from a phenomenological point of view, we anticipate that the growth should be positive for $\tau_E\lesssim\tau_H$.
In order to circumvent this issue we cut off contributions from modes with $\chi\tau_E(q)>\tau_H$.
However, for sufficiently large $\tau_D\sim\tau_H$ this might not be sufficient.
Note that for nonhelical turbulence this is not an issue even for $\tau_D\sim\tau_H$ and consequently we anticipate that a proper treatment of the eddy turnover time for helical turbulence might suffice in properly resolving the issue.
In figure \ref{heliplo} we show the GW spectrum from MHD turbulence with maximal magnetic helicity for different integration times and the case $\chi=0$ and $\chi=2$.
As can be seen for $\chi=0$ the energy spectrum i.e. decreases at later times at frequencies below the peak frequencies.
For $\chi=2$ an energy decrease is still observable and quite significant, but in the high frequency tail, such a criterion is sufficient to make certain that the energy rate remains positive.
Nonetheless, in general a simple cutoff will not suffice, in particular not in the intermediary range.
Generally, as argued before we do not expect this to be a significant problem for $\tau_D\ll\tau_H$.
Moreover for $\chi=2$ one finds a double peak spectrum emerging and in generally we expect this to be a clear artifact of the modelling of the decorrelation for helical scenarios.
We investigate the impact of helicity further below.

Now, we investigate the dependence of the spectrum on the initial integral scale of the turbulence $L_*=2H_*/(\beta v_w)$ and the dependence on the initial turbulent 
energy parameter $\Omega_{t,*}=\alpha\kappa_v$.
 \begin{figure}[htbp]
 \centering
 \begin{minipage}{0.45\textwidth}
 \includegraphics[scale=0.55]{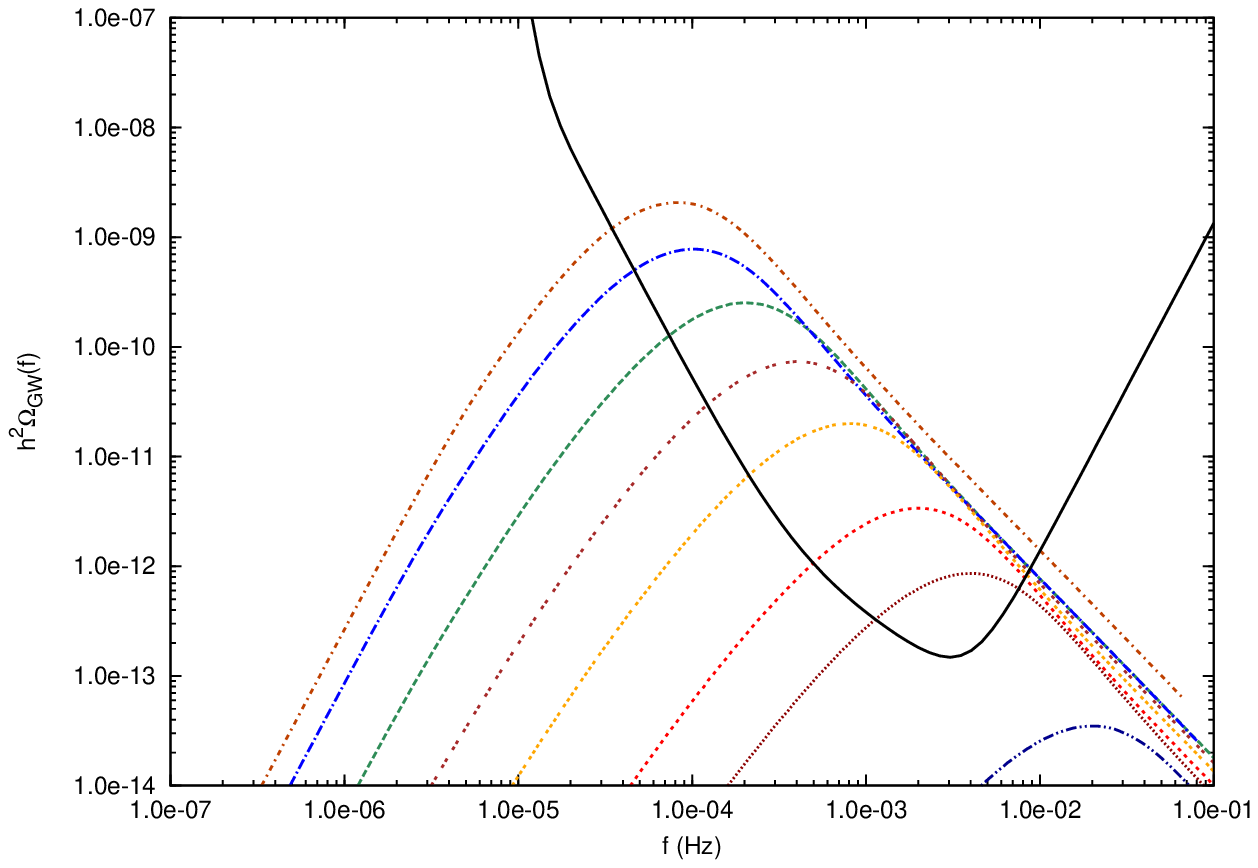}
 \end{minipage}
 \hfill
 \begin{minipage}{0.45\textwidth}
 \includegraphics[scale=0.55]{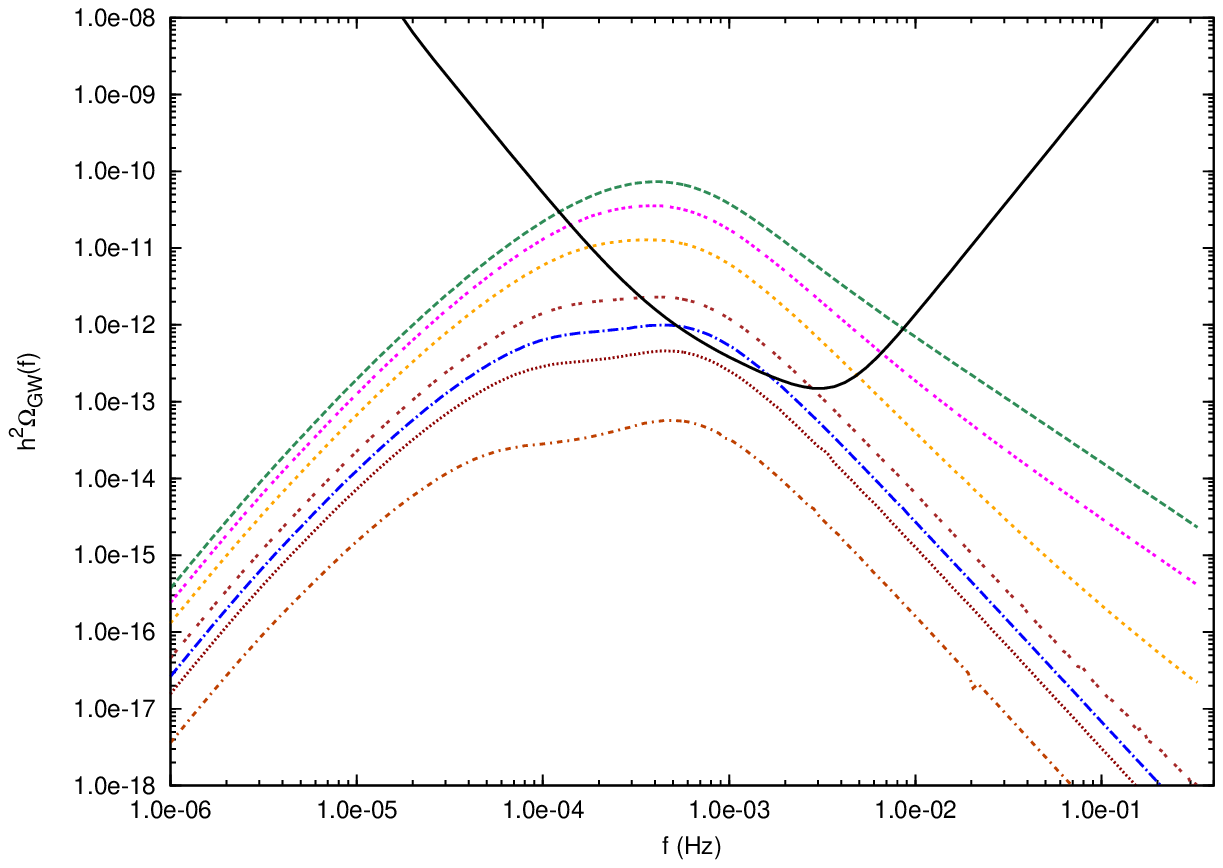}
 \end{minipage}
 \caption{On the left panel, the dependence of the spectrum on the initial value $L_*$ is shown, where $\Omega_{t,*}/\Omega_r=0.2$ ($\alpha\sim0.7$) and $T_*=100$GeV have been chosen.
 From top to bottom the lines correspond to $L_* H_*=0.4,0.2,0.1,0.05,0.025,0.01,0.005,0.001$ (dark-orange, blue, green, brown, orange, red, dark-red, dark-blue).
 On the right panel, the dependence on $\Omega_{t,*}$ is investigated for $L_* H_*=0.1$ ($\beta/H_*\sim20$), where from top to bottom the lines correspond to $\Omega_{t,*}/\Omega_r=0.2,0.15,0.1,0.05,0.035,0.025,0.01$
 (green, magenta, orange-yellow, brown, blue, dark-red, dark-orange).
 In both figures and for all scenarios we set $\chi=0$.}
 \label{standard-scen}
 \end{figure}
As can be seen in figure \ref{standard-scen} the total power scales as $L_*^{-2}$.
The shape of the spectrum itself does not change when $L_*$ changes.
There is a clear deviation towards large values of $L_* H_*$ in the shape of the spectrum.
Moreover, the shape of the spectrum itself strongly depends on $\Omega_{t,*}$, as can be seen in figure (\ref{standard-scen}).
In particular for small $\Omega_{t,*}/\Omega_r$, conforming with the causality constraint $\tau_D\lesssim \tau_H$, corresponding to $\sqrt{\Omega_{t,*}/\Omega_r}\gtrsim L_* H_*$, we find that the spectral 
shape in the large frequency region follows a $f^{-8/3}$ scaling. 
Also, it appears that at very large frequencies a $f^{-5/3}$ scaling can be observed which requires a Kolmogorov turbulence spectrum and a sufficiently large inertial range.
On the other hand, a $f^{-5/3}$ scaling is observed for sufficiently large $\Omega_{t,*}/\Omega_r$.
Furthermore, the scaling around the peak of the spectrum also changes with decreasing $\Omega_{t,*}$.
Therefore, we find that the spectral index of the large frequency tail of the gravitational wave spectrum explicitly depends on $\Omega_{t,*}$ and overall that this range is described 
by two different power laws.
In the previous estimate for the shape of the spectrum, we generally assumed that the relevant timescales have to be much smaller than the Hubble time, yet for smaller values 
of $\Omega_{t,*}$ the impact of the cosine becomes quite relevant.
In general the overall amplitude of the spectrum scales differently with $\Omega_{t,*}$ at small and at large scales.
At small frequencies we find a dependence that is approximately compatible with $\Omega_{t,*}^{3/2}$, whereas at large frequencies the dependence on $\Omega_{t,*}$ is more complicated due to the fact that
two power laws appears with a range that depends on $\Omega_{t,*}$ itself, e.g. a simple power law index scaling is not possible for that range, yet a scaling $\propto\Omega_{t,*}^3$ for large $\Omega_{t,*}$
can be estimated, whereas for smaller $\Omega_{t,*}$ the scaling would be far steeper.
Here we do not attempt to give a simple formula, due to the more complicated dependencies on $\Omega_{t,*}$ and hence on $\alpha$.

Next, a similar analysis for the maximal helical scenario is presented and we study the dependence of the GW spectrum on $\Omega_{t,*}$ and $L_*$, while we fix $\chi=2$. 
\begin{figure}[htbp]
\centering
\begin{minipage}{0.45\textwidth}
\includegraphics[scale=0.55]{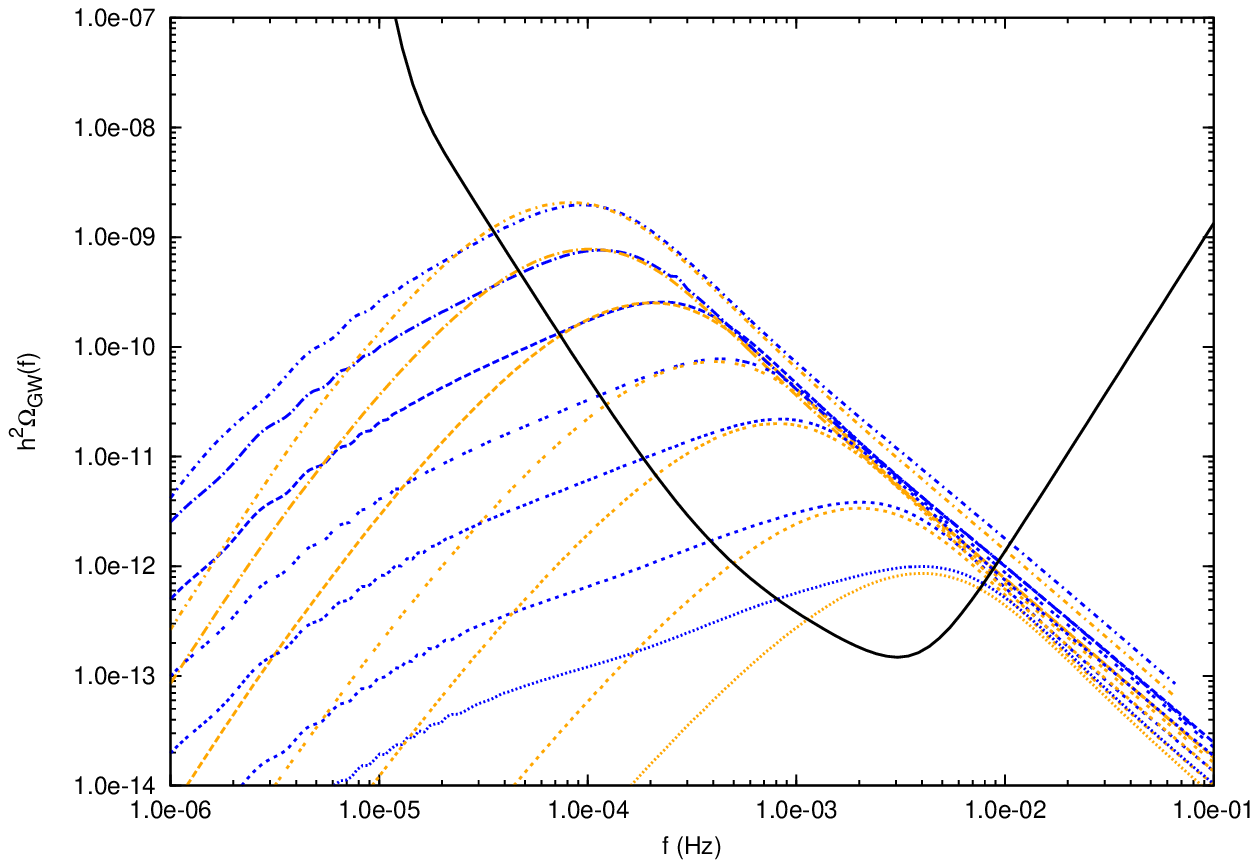}
\end{minipage}
\hfill
\begin{minipage}{0.45\textwidth}
\includegraphics[scale=0.55]{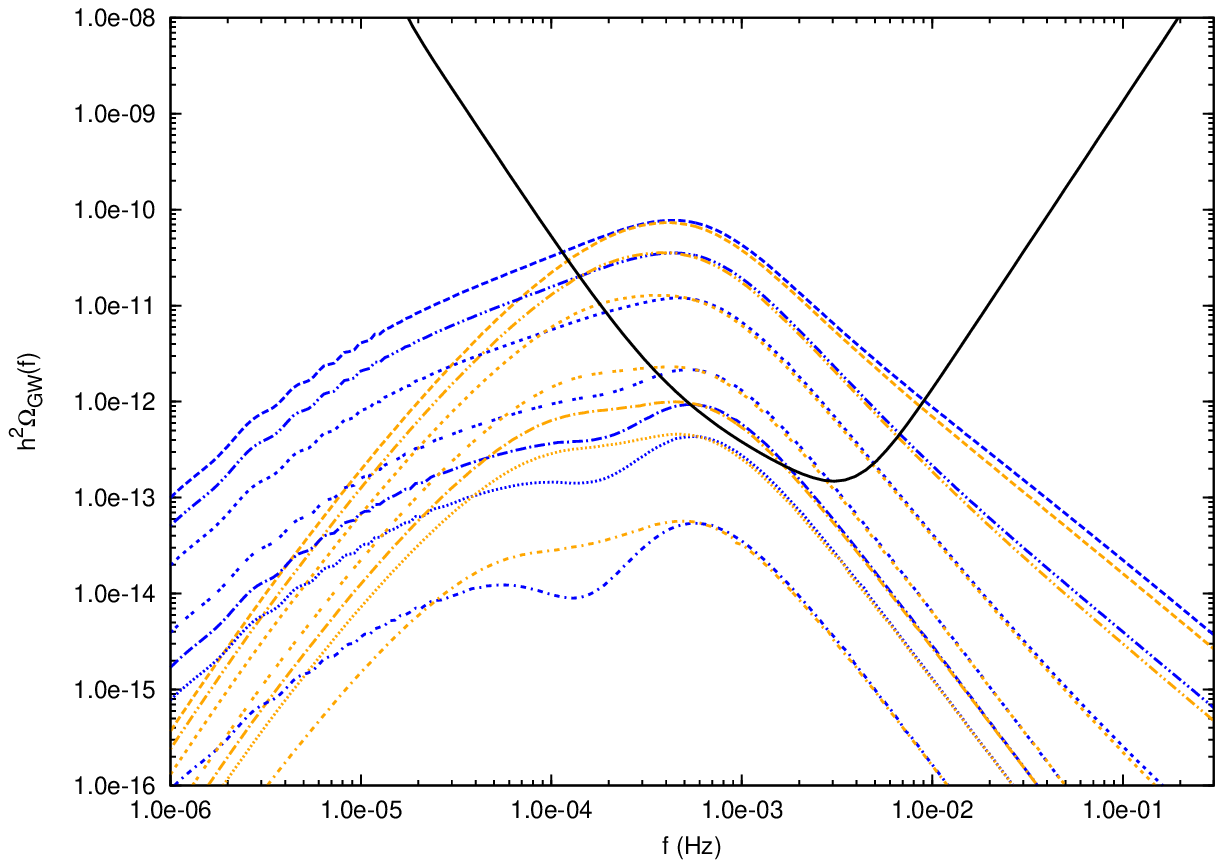}
\end{minipage}
\caption{On the left panel, the dependence of the spectrum on the initial value $L_*$ is shown, where $\Omega_{t,*}/\Omega_r=0.2$ ($\alpha\sim0.7$) and $T_*=100$GeV have been chosen.
From top to bottom the lines correspond to $L_* H_*=0.4,0.2,0.1,0.05,0.025,0.01,0.005$, where blue lines denote the helical case, while orange lines denote the non-helical case.
On the right panel, the dependence on $\Omega_{t,*}$ is investigated for $L_* H_*=0.1$ ($\beta/H_*\sim20$).
We show again both the helcial (blue) and nonhelical (orange) scenario.
From top to bottom the lines correspond to $\Omega_{t,*}/\Omega_r=0.2,0.15,0.1,0.05,0.035,0.025,0.01$.
In both plots we have fixed $\chi=2$ (only modes with $\chi\tau_E(k)<\tau_H$ contribute).}
\label{helical-scen}
\end{figure}
As can be seen in figure (\ref{helical-scen}) the $L_*$ and $\Omega_{t,*}$ dependence is only slightly more complicated.
Overall the helical spectra have a flatter low-frequency spectrum, which becomes slightly more pronounced for lower $L_*$.
The range of the flatter spectrum is directly dependent on the integration time, here we integrated only over one magnitude in scale factor and hence for helical 
scenarios gravitational waves will generally be seeded at later times, however for the particular examples here, a longer integration is not needed since it is well 
out of the range of planned detectibility.
However helical turbulence could produce signals that record a longer evolution of the universe and are hence of interest.
At the high frequency tail, helicity leads in general to minor changes.
Overall the dependence is comparable to the nonhelical case.
Regarding the $\Omega_{t,*}$ dependence, we find for large $\Omega_{t,*}$, that the spectrum shows an increase in energy at all scales and overall the dependence is also comparable to that of the nonhelical case.
In general, we see for most studied scenarios, an increase in energy at small frequencies, a decrease at intermediary frequencies and an increase around the peak and at least partially in the large frequency tail.
However, we suspect that the decrease at intermediary frequencies is likely an artifact of the insufficient decorrelation function for helical scenarios, as discussed before.
Since $\tau_D\propto L_I/\sqrt{\Omega_t}$, we note that scenarios with small $L_I$ overall should be least bothered by the suspected artifacts, whereas scenarios 
with smaller $\Omega_t$ should be more bothered by these artifacts as one can seen in the double peak features in figure (\ref{helical-scen}).

Lastly, we focus now on a very important aspect i.e. the dependence of the GW spectrum on a non-zero fraction of dilatational fluid motion for a scenario with $\Omega_{t,*}/\Omega_r=0.13$ ($\alpha=0.5$), 
$L_*=0.2$ ($\beta/H_*\sim10$) and we set for the build time $\tau_b=1/\beta$.
\begin{figure}[htbp]
\centering
\begin{minipage}{0.45\textwidth}
\includegraphics[scale=0.55]{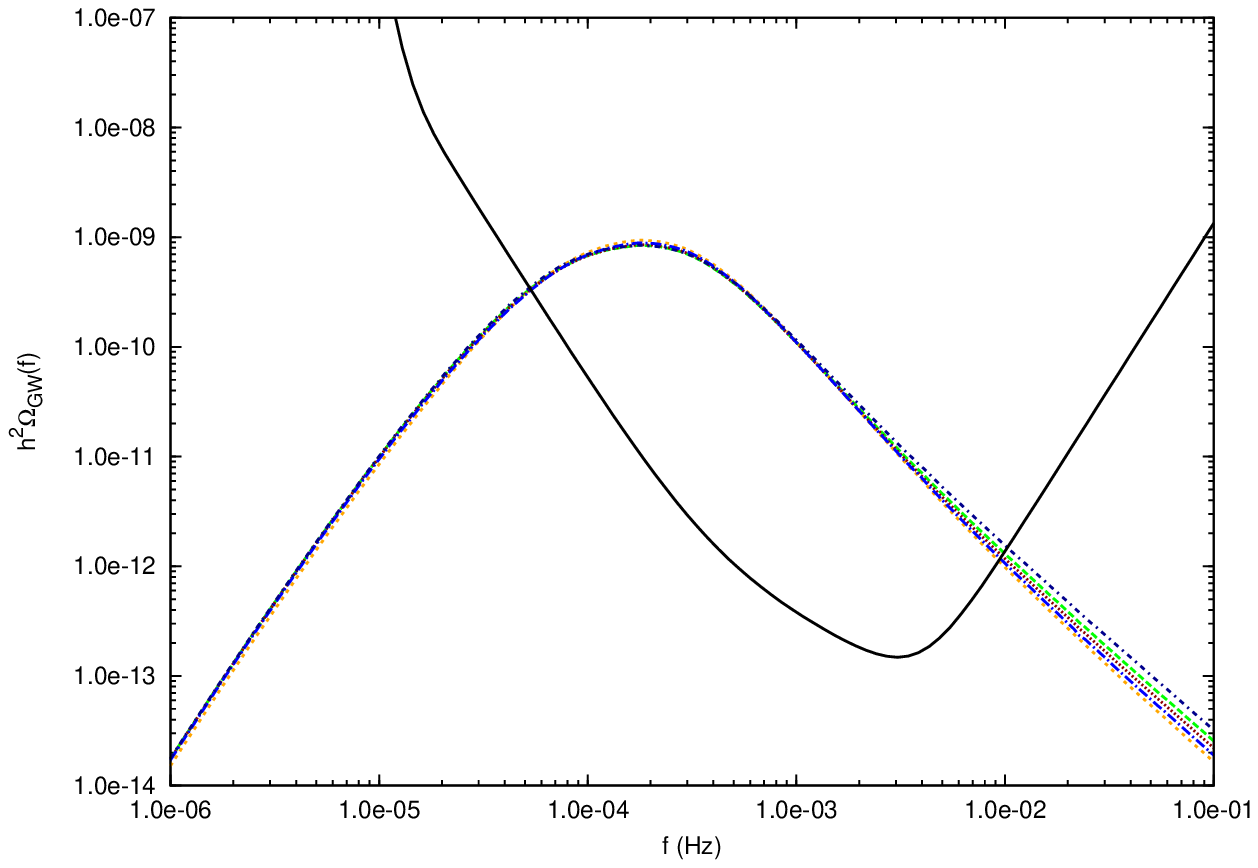}
\end{minipage}
\hfill
\begin{minipage}{0.45\textwidth}
\includegraphics[scale=0.55]{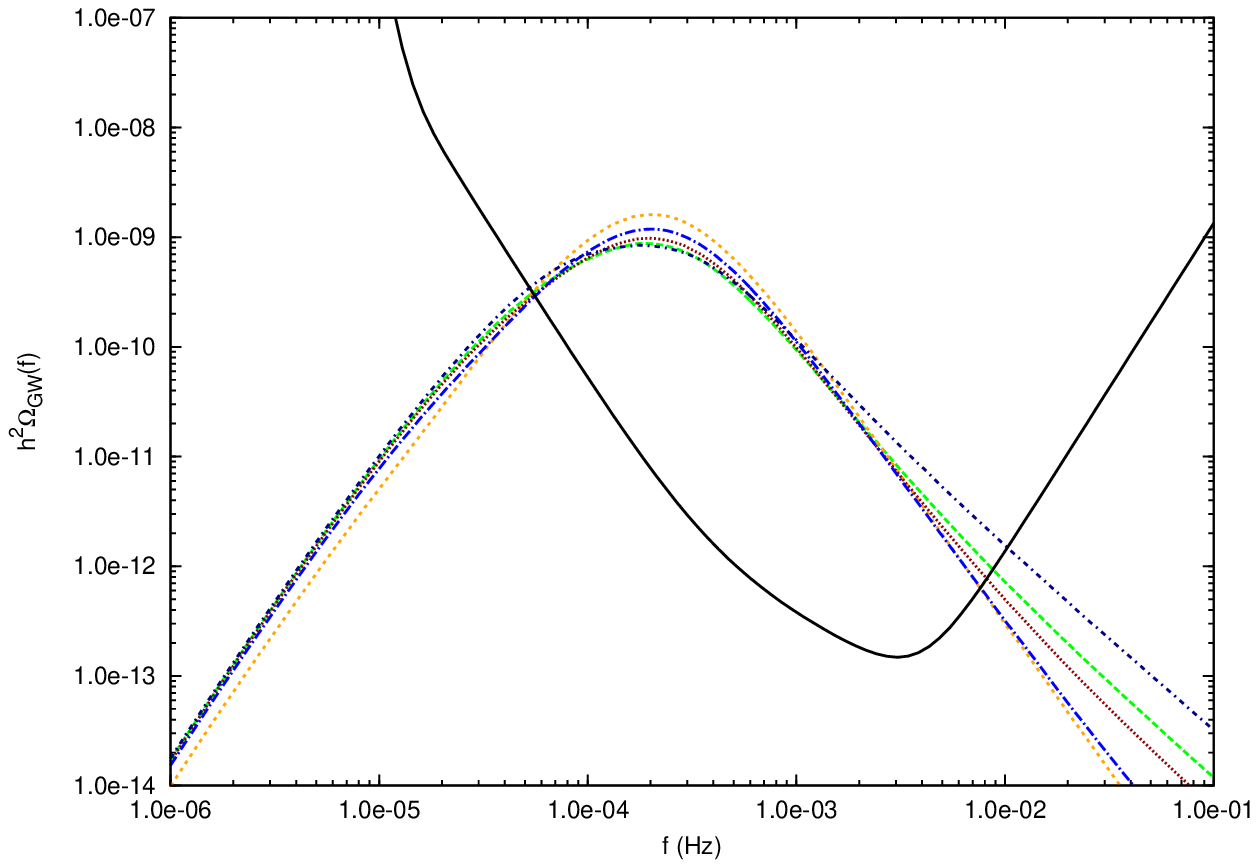}
\end{minipage}
\\
\begin{minipage}{0.45\textwidth}
\includegraphics[scale=0.55]{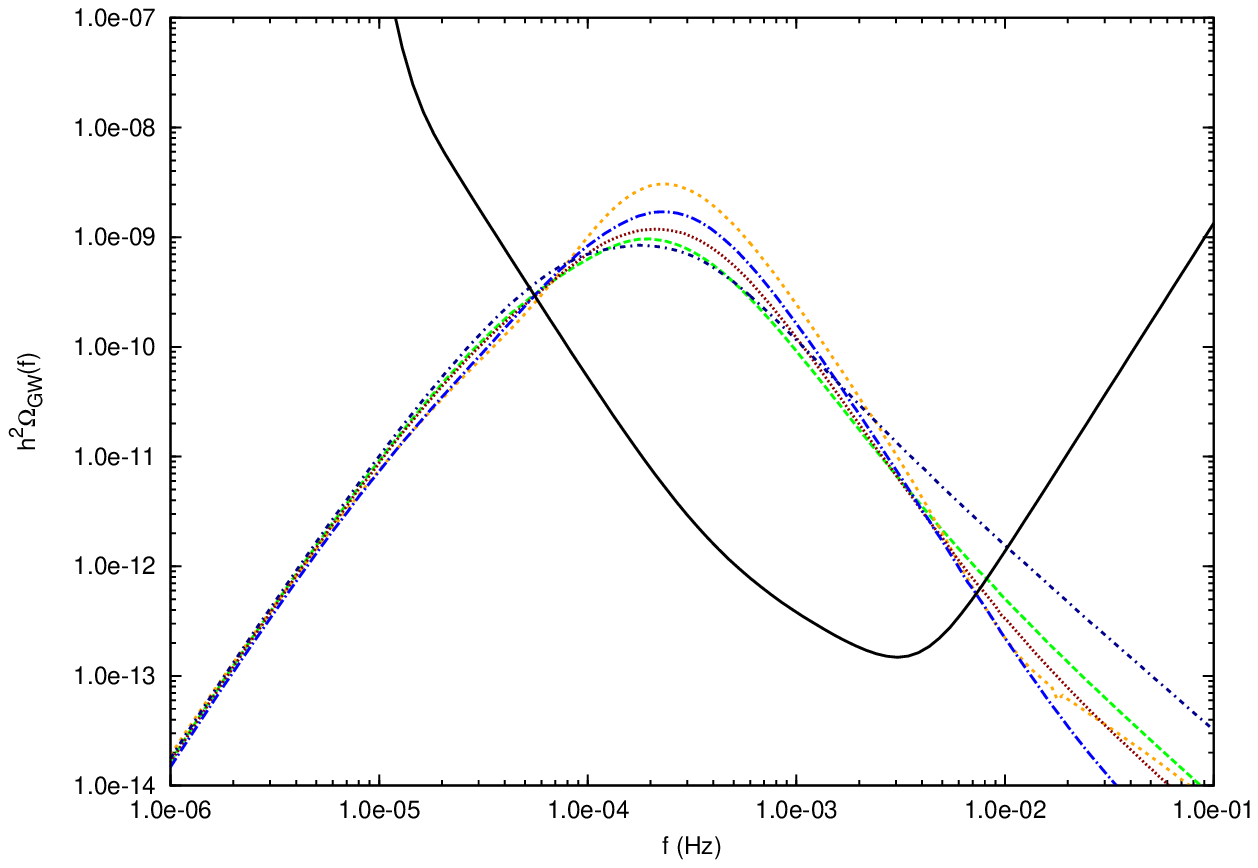}
\end{minipage}
\hfill
\begin{minipage}{0.45\textwidth}
\includegraphics[scale=0.55]{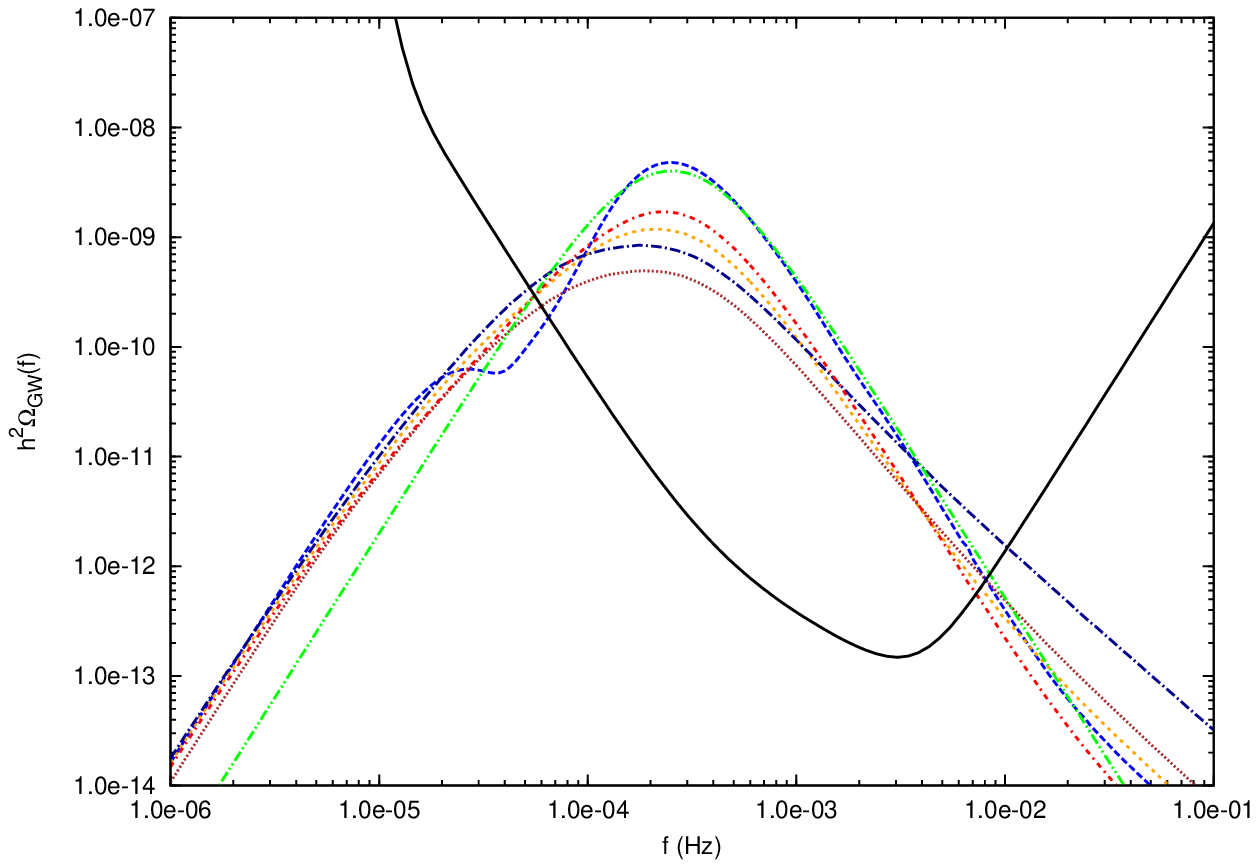}
\end{minipage}
\caption{The gravitational wave power spectrum predicted by different values for the initial fraction of dilatational modes $f_D=0.1$ in the top left, $f_D=0.5$ in the top right and $f_D=0.9$ in the bottom left panel.
The different lines correspond to the different scenarios: model A (\ref{modA}) (green, dashed), model B (\ref{modB}) (orange, thick-dotted), model C$_1$ (\ref{modC}) 
(dark-red, thin-dotted) and model C$_2$ (blue, dot-dashed) for the temporal evolution of $f_D$. Each model is also compared with the case $f_D=0$ (dark-blue, dot-dashed).
In the bottom right panel, we show the cases $f_D=0$ (dark-blue, dot-dashed) with $\tau_b=\beta^{-1}$ (dot-dashed, dark blue) and $\tau_b=\tau_D$ (thin-dotted, brown), $f_D=0.9$ with model C$_1$ and $\tau_b=\beta^{-1}$ (blue, dashed), 
$f_D=0.9$ (orange, thick-dotted) with model C$_2$ and $\tau_b=\beta^{-1}$, $f_D=0.999$ (red, dot-dashed) with model B and $\tau_b=\beta^{-1}$, 
and extrapolated fitted DNS results (green, double dot-dashed) from \protect\citeasnoun{Caprini2018} via \protect\cite{HindmarshHuber2015}.}
\label{dilap}
\end{figure}

In figure (\ref{dilap}) one sees that for a small fraction of dilatational fluid motion ($f_D=0.1$), the shape of the high frequency tail is slightly affected and the GW spectrum at those scales is slightly smaller.
In particular the power-law index of the high frequency spectrum becomes closer to $-2$. 
Overall the different models for the evolution of the fraction of the dilatational modes discussed here, model A (\ref{modA}), B (\ref{modB}) and C (\ref{modC}), i.e. a fraction $f_D$ decreasing on a timescale $\tau_D$, a constant $f_D$ and 
an initially constant $f_D$ that after some time starts to decay, differ only slightly.
For $f_D=0.5$ the picture overall is more diverse.
At the peak frequency the scenario of constant $f_D$ leads to an amplification of a factor of 2.
At higher frequencies the deviation becomes more pronounced and also the overall shape of the spectrum is more steep.
There is a shift from a $f^{-5/3}$ to a $f^{-2}$ high frequency tail for slightly smaller solenoidal energy densities, due to the dilatational contribution becoming dominant,
while still driven by the decorrelation due to the solenoidal modes.
Next, for $f_D=0.9$ one sees a stronger distinction between the different models, i.e. for constant $f_D$ the difference has grown to a factor 4 in comparison to the purely solenoidal case and the spectrum 
has an $f^{-3}$ scaling which towards higher frequencies evolves into an $f^{-2}$ tail.
Whereas, the other dilatational scenarios develop a $f^{-2}$ high frequency tail at frequencies above the peak.
In the lower right panel of figure (\ref{dilap}), different scenarios are compared and one sees that the two extreme cases i.e. the purely dilatational flow and the purely solenoidal flow differ by up to an order of 
magnitude at the peak.
At high frequencies the scaling for the purely incompressible case is $f^{-5/3}$ and $f^{-3}$ for the purely compressible flow.
The latter case is also in accordance with direct numerical simulations extrapolated to the Hubble time from \citeasnoun{Caprini2018} based on \citeasnoun{HindmarshHuber2015}.
One important feature appearing is at frequencies below the peak frequency, where the spectrum overall differs considerable from the extrapolated results by \cite{HindmarshHuber2015}.
It is not clear what causes this behavior, but it might be related to overly long correlation times in the calculation.
Overall, we generally expect for a strongly first order phase transition, as indicated by \cite{Hindmarsh2017} that solenoidal modes dominate
corresponding to values for $f_D$ between $\sim0.1$ and $\sim0.5$.
Therefore we anticipate a $\sim f^{-2}$ high frequency tail for such phase transitions.
For weakly first order phase transitions, we anticipate that the $f^{-3}$ power law shown by \cite{HindmarshHuber2015} should be observed, although the amplitude might nontheless be smaller as it depends on 
the maximal integration time.
One other important factor is that, in previous analysis the comparison between the soundwave and "turbulent" scenario was based on different build up-time $\tau_b=\tau_D$ for the turbulent case and $\tau_b=\beta^{-1}$
for the soundwave case, yet as can be seen in the right bottom panel in figure (\ref{dilap}) this in by itself is responsible for around a factor of 2 of the difference between these two cases.

\section{Summary and Outlook}
We have performed semi-analytical calculations of the GW power spectrum $\Omega_{GW}(f)$ due to compressible MHD turbulence and have shown that the sweeping effect leads to a decrease of its amplitude compared
to previous calculations for incompressible turbulence.
Additionally, we have discussed the impact of magnetic helicity and find that it will strongly impact the shape of the spectrum at low frequencies.
However, we also note that there are still difficulties in the calculation of the GW spectrum from helical turbulence. 
For purely incompressible turbulence we also observe that $\Omega_{GW}(f)$ is proportional to $L_*^2$ and $\Omega_{t,*}^{3/2}$ below the peak frequency and scales in a more complicated fashion with $\Omega_{t,*}$
above.
In particular a power law scaling $f^{-5/3}$ for large $\Omega_{t,*}$ and $f^{-8/3}$ for small $\Omega_{t,*}$ is observed.
In general due to the less efficient production of GWs by MHD turbulence, resulting constraints on magnetic fields produced at a first order EWPT will be weakened within our approach.

To our knowledge for the first time we have studied, using different toy models, the impact of solenoidal modes on the gravitational wave signal from dilatational modes e.g. sound waves.
We find for a strong first order phase transitions a scaling with $f^{-2}$ for the high frequency tail of $\Omega_{GW}$, due to the sweeping effect of solenoidal modes on dilatational modes.
Our semi-analytical approach is an alternative to full blown numerical simulations, which in general are difficult to extrapolate to sufficiently large timescales.
In general we expect that a direct extrapolation of the GW sourcing by sound-waves leads to an overestimate of the gravitational wave energy density, 
since even a minor fraction of solenoidal modes ($f_S\sim0.1$) will greatly 
reduce the GW production efficiency of sound-waves over a Hubble time for phase transition scenarios with a causal eddy turnover time.

One key aspect, which requires more clarification is the impact of magnetic fields on unequal time correlations.
Further, the formation of shocks can modify the picture for dilatational dominated turbulence, in cases where the eddy turnover time is bigger than the Hubble time.
We expect that the gravitational wave signal predicted by a more detailed treatment of the evolution of dilatational modes will be contained within the range of results obtained from the toy models studied here.

Magnetic fields itself also act as a source of rotational motion and a feedback loop of magnetic field and rotational motion (dynamo) might lead to a more efficient transformation of dilatational to 
solenoidal modes, which could be relevant for intermediary strong phase transitions.
Also the precise dependence of $\Omega_{GW}(f)$ on $\Omega_{t,*}$ and $L_*$, especially with regard to compressible turbulence,  will require further studies.
Further useful information might be contained in the polarization spectrum of GWs, which was not considered here.

\section*{Acknowledgments}
We thank Andrej Dundovic, Andrey Saveliev, Petar Pavlovic and Pranjal Trivedi for helpful discussions.
Further, we thank Geraldine Servant for helpful comments.
This work was supported by the Deutsche Forschungsgemeinschaft through the Collaborative Research Center SFB 676 "Particles, Strings and the Early Universe".

\appendix

\section{Calculation of Structure Coefficients and Time Ordering}
Here we give a brief overview of some steps that are required for the calculation of (\ref{gwpowspec}).
\subsection{Product of sine and cosine functions}
The product of (\ref{changestrain}) with itself , which is the gravitational wave energy, involves a product of cosine and sine functions
\begin{align}
 F_{\rm RD}(k,k',\tau,\tau',\tau'')=&\cos(k(\tau-\tau'))\cos(k'(\tau-\tau''))+\frac{1}{kk'\tau^2}\sin(k(\tau-\tau'))
 \sin(k'(\tau-\tau''))\\
 &-\frac{1}{k'\tau}\cos(k(\tau-\tau'))\sin(k'(\tau-\tau''))-\frac{1}{k\tau}\cos(k'(\tau-\tau''))\sin(k(\tau-\tau''))\nonumber.
\end{align}
Here we are primarily interested in modes that are well within the horizon $k\tau\gg 1$ (causal modes).
Thus, we neglect all contributions of $\mathcal{O}((k\tau)^{-1})$ in $F_{\rm RD}$, which reduces to
\begin{equation}
F_{\rm RD}(k,k',\tau,\tau',\tau'')\approx\cos(k(\tau-\tau'))\cos(k'(\tau-\tau'')).
\end{equation}
Besides, the function $F_{\rm RD}$ varies quite drastically over these causal modes, since $2\pi/(k\tau)\ll 1$.
Averaging over these modes for $k=k'$ leads to
\begin{equation}
  F_{\rm RD}(k,k,\tau',\tau',\tau'')\approx\cos(k(\tau'-\tau''))/2.
\end{equation}
Now, $F_{\rm RD}$ no longer depends on $\tau$.

\subsection{Structural coefficients}
First, we evaluate the correlator $\ev{\pi_{ij}(\vec{k},\tau')\pi_{ij}^{*}(\vec{k}',\tau'')}$
 \begin{align}
\ev{\pi_{ij}^T(\vec{k},\tau')\pi_{ij}^{T*}(\vec{k}',\tau'')}=&P^2_{ijab}(\vec{k})P^2_{ijcd}(\vec{k'})\frac{(\rho+p)^2}{(2\pi)^6}
\int{\rm d}^3\vec{q}
\int{\rm d}^3\vec{q}'\Bigl[\ev{b_a(\vec{q},\tau')b_b(\vec{p},\tau')b_c^*(\vec{q'},\tau'')b_d^*(\vec{p'},\tau'')}\\
&+\ev{v_a(\vec{q},\tau')v_b(\vec{p},\tau')v_c^*(\vec{q'},\tau'')v_d^*(\vec{p'},\tau'')}\Bigr],\nonumber
\end{align}
where $\vec{p}'=\vec{q}'-\vec{k}'$.
Next we focus on the four point velocity correlation
\begin{align}
\ev{v_a(\vec{q},\tau')v_b(\vec{p},\tau')v_c^*(\vec{q'},\tau'')v_d^*(\vec{p'},\tau'')}=&\ev{v_a(\vec{q},\tau')v_c^*(\vec{q'},\tau'')}\ev{v_b(\vec{p},\tau')v_d^*(\vec{p'},\tau'')}+\\
&\ev{v_a(\vec{q},\tau')v_c^*(\vec{p'},\tau'')}\ev{v_b(\vec{p},\tau')v_d^*(\vec{q'},\tau'')},\nonumber
\end{align}
where correlations of the type $\ev{v_a(\vec{q},\tau')v_b(\vec{p},\tau')}$ haven been neglected, since these imply $\vec{p}=-\vec{q}$ and this requires $\vec{k}=0$.
The correlation function imply $\vec{k}=\vec{k}'$ for the other terms.
Then, the system is reduced to two point-functions and can now be evaluated using (\ref{uneqtimecorvel}) for solenoidal velocity fluctuations, (\ref{uneqtimecor}) for magnetic fluctuations and (\ref{sweptwave}) for dilatational velocity fluctuations.

For different combinations of modes the structural coefficients for solenoidal and dilatational modes in (\ref{spmcoff}) and (\ref{dcoff}), corresponding to the product of the 
spectral tensors of the different two point functions with the quadratic projectors, are given by
\begin{equation}
 2P^2_{ijab}(k)P^2_{ijcd}(k)P_{ac}(q)P_{bd}(p)=2P^2_{abcd}(k)P_{ac}(q)P_{bd}(p)=1+2\left[(\hat{k}\cdot\hat{q})^2+(\hat{k}\cdot\hat{p})^2\right]+(\hat{q}\cdot\hat{k})^2(\hat{k}\cdot\hat{p})^2 \label{corppro}
\end{equation}
for terms that involve only solenoidal fluctuations e.g. $E_p^S E_q^S$.
The factor 2 is due to the symmetry in $p$ and $q$ in Wick's theorem.
This coefficient differs slightly from the one calculated in e.g. \cite{Caprini2002}.
Therefore, we give a step by step  calculation of (\ref{corppro})
\begin{align}
 2P^2_{abcd}(k)P_{ac}(q)P_{bd}(p)=&2P_{ac}(k)P_{ac}(q)P_{bd}(k)P_{bd}(p)-P_{ab}(k)P_{ac}(q)P_{cd}(k)P_{bd}(p)\nonumber\\
 =&2\left(1+(\hat{k}\cdot\hat{p})^2\right)\left(1+(\hat{k}\cdot\hat{q})^2\right)\nonumber\\
 &-P_{ab}(k)P_{cd}(k)\left(\delta_{ac}\delta_{bd}-\delta_{ac}\frac{p_bp_d}{p^2}-\delta_{bd}\frac{q_aq_c}{q^2}+\frac{p_bp_d}{p^2}\frac{q_aq_c}{q^2}\right),
\end{align}
where we used $P_{ij}(q)P_{ij}(k)=1+(\hat{k}\cdot\hat{q})^2$.
We now evaluate the other products of the projectors
\begin{equation}
 -P_{ab}(k)P_{cd}(k)\delta_{ac}\delta_{bd}=-P_{ac}(k)\delta_{ac}=-2,
\end{equation}
where we used $P_{ab}(k)P_{bc}(k)=P_{ac}(k)$.
Moreover we have the terms
\begin{equation}
 P_{ab}(k)P_{cd}(k)\delta_{bd}\frac{q_aq_c}{q^2}=P_{ac}(k)\frac{q_aq_c}{q^2}=1-(\hat{k}\cdot\hat{q})^2
\end{equation}
and analogously
\begin{equation}
 P_{ab}(k)P_{cd}(k)\delta_{ac}\frac{p_bp_d}{p^2}=P^{bd}(k)\frac{p_bp_d}{p^2}=1-(\hat{k}\cdot\hat{p})^2.
\end{equation}
For the last term, we use $P_{ab}(k)k_a=0$ and $p_a=k_a-q_a$ to get
\begin{equation}
 -P_{ab}(k)P_{cd}(k)\frac{p_bp_d}{p^2}\frac{q_aq_c}{q^2}=-P_{ab}(k)\frac{p_ap_b}{p^2}P_{cd}(k)\frac{q_cq_d}{q^2}=-\left(1-(\hat{k}\cdot\hat{q})^2\right)\left(1-(\hat{k}\cdot\hat{p})^2\right).
\end{equation}
These four terms together give $-1-(\hat{k}\cdot\hat{q})^2(\hat{k}\cdot\hat{p})^2$ and hence we have
\begin{align}
 2P^2_{abcd}(k)P_{ac}(q)P_{bd}(p)&=2\left(1+(\hat{k}\cdot\hat{p})^2\right)\left(1+(\hat{k}\cdot\hat{q})^2\right)-1-(\hat{k}\cdot\hat{q})^2(\hat{k}\cdot\hat{p})^2\nonumber\\
 &=1+2\left[(\hat{k}\cdot\hat{q})^2+(\hat{k}\cdot\hat{p})^2\right]+(\hat{q}\cdot\hat{k})^2(\hat{k}\cdot\hat{p})^2.
\end{align}

For the other terms we only give the results.
Next, for terms that involve only dilatational fluctuations, one finds
\begin{equation}
 2P^2_{abcd}(k)4\frac{q_aq_c}{q^2}\frac{p_bp_d}{p^2}=4\left(1-\left[(\hat{k}\cdot\hat{q})^2+(\hat{k}\cdot\hat{p})^2\right]+(\hat{q}\cdot\hat{k})^2(\hat{k}\cdot\hat{p})^2\right).
\end{equation}
Moreover one also has different coefficients for terms involving $E_q^D E_p^S$
\begin{equation}
2P^2_{abcd}(k)2\frac{q_aq_c}{q^2}P_{bd}(p)=6(\hat{k}\cdot\hat{p})^2\left(1-(\hat{q}\cdot\hat{k})^2\right)
\end{equation}
and 
\begin{equation}
2P^2_{abcd}(k)2\frac{p_ap_c}{p^2}P_{bd}(q)=6(\hat{k}\cdot\hat{q})^2\left(1-(\hat{p}\cdot\hat{k})^2\right)
\end{equation}
for terms like $E_p^D E_q^S$.
Lastly for the helical terms e.g. $h_{V}(q)h_{V}(p)$ one finds
\begin{equation}
2P^2_{abcd}(k)\epsilon_{acn}\frac{q_n}{q}\epsilon_{bdm}\frac{p_m}{p}=2(\hat{k}\cdot\hat{q})(\hat{k}\cdot\hat{p}).
\end{equation}

\subsection{Time ordering}
Now, we combine the above results and we find for the gravitational wave energy density
\begin{align}
 \rho_G(\vec{x},\tau)=& \frac{4\pi}{(4\pi)^2} G_{\rm N}H_0^{-2}\Omega_r f_g(\rho+p)^2\int{\rm d}^3\vec{k}\int{\rm d}^3\vec{q}\int_{\tau_0}^\tau {\rm d}\tau'\
 \int_{\tau_0}^\tau {\rm d}\tau'' \frac{1}{q^3 p^3 \tau\tau'}\cos(k(\tau'-\tau'')) \nonumber \\ 
 &\times f_{\rm RSA}(\tau',\tau'',q)f_{\rm RSA}(\tau',\tau'',p)\Bigl[E_t^2(q,p,\tau')S^+(k,q,p)+4H_t^2(q,p,\tau')(\hat{k}\cdot\hat{q})(\hat{k}\cdot\hat{p})+\nonumber\\
& S^-(k,q,p)\Bigl(4E_D(q,\tau')E_D(p,\tau')\cos\left[qc_s(\tau'-\tau'')\right]\cos\left[pc_s(\tau'-\tau'')\right]\Bigr)+\nonumber\\
 &6D(k,p,q)E_S(q,\tau')E_D(p,\tau')\cos\left[pc_s(\tau'-\tau'')\right]+\nonumber\\
 & 6D(k,q,p)E_S(p,\tau')E_D(q,\tau')\cos\left[qc_s(\tau'-\tau'')\right]\Bigr].  \label{gwenunord}
 \end{align}

In the next step we bring (\ref{gwenunord}) into a time ordered form by differentiating and reintegrating.
The time derivative of the gravitational wave energy is
 \begin{align}
 \partial_\tau\rho_G(\vec{x},\tau)\approx& \frac{8\pi}{(4\pi)^2} G_{\rm N}H_0^{-2} f_g\Omega_r(\rho+p)^2\int{\rm d}^3\vec{k}\int{\rm d}^3\vec{q}
 \int_{\tau_0}^\tau {\rm d}\tau'\
  \frac{1}{q^3 p^3 \tau\tau'} \cos(k(\tau-\tau')) \nonumber \\
 &\times f_{\rm RSA}(\tau,\tau',q)f_{\rm RSA}(\tau,\tau',p)\Bigl[E_t^2(q,p,\tau')S^+(k,q,p)+4H_t^2(q,p,\tau')(\hat{k}\cdot\hat{q})(\hat{k}\cdot\hat{p})+\nonumber\\
& S^-(k,q,p)\Bigl(4E_D(q,\tau')E_D(p,\tau')\cos\left[qc_s(\tau'-\tau'')\right]\cos\left[pc_s(\tau'-\tau'')\right]\Bigr)+\nonumber\\
 &6D(k,p,q)E_S(q,\tau')E_D(p,\tau')\cos\left[pc_s(\tau'-\tau'')\right]+\nonumber\\
 & 6D(k,q,p)E_S(p,\tau')E_D(q,\tau')\cos\left[qc_s(\tau'-\tau'')\right]\Bigr]. \label{gwenordder}
 \end{align}
Note that the energies are now evaluated at $\tau$ (Markov-form) and no longer at intermediary times for the rate of change of the 
GW-energy.
Lastly, we reintegrate this expression again over all times $\tau$ and find
 \begin{align}
 \rho_G(\vec{x},\tau)\approx& \frac{8\pi}{(4\pi)^2} G_{\rm N}H_0^{-2}\Omega_r f_g(\rho+p)^2\int{\rm d}^3\vec{k}\int{\rm d}^3\vec{q}
 \int_{\tau_0}^\tau {\rm d}\tau'\ \int_{\tau_0}^{\tau'} {\rm d}\tau''\
  \frac{1}{q^3 p^3\tau'\tau''} \cos(k(\tau'-\tau'')) \nonumber \\
 &\times f_{\rm RSA}(\tau',\tau'',q)f_{\rm RSA}(\tau',\tau'',p)\Bigl[E_t^2(q,p,\tau')S^+(k,q,p)+4H_t^2(q,p,\tau')(\hat{k}\cdot\hat{q})(\hat{k}\cdot\hat{p})+\nonumber\\
& S^-(k,q,p)\Bigl(4E_D(q,\tau')E_D(p,\tau')\cos\left[qc_s(\tau'-\tau'')\right]\cos\left[pc_s(\tau'-\tau'')\right]\Bigr)+\nonumber\\
 &6D(k,p,q)E_S(q,\tau')E_D(p,\tau')\cos\left[pc_s(\tau'-\tau'')\right]+\nonumber\\
 & 6D(k,q,p)E_S(p,\tau')E_D(q,\tau')\cos\left[qc_s(\tau'-\tau'')\right]\Bigr].  
 \end{align}

\section*{References}
\bibliography{GW_MHD}
\bibliographystyle{jphysicsB}

\end{document}